\documentclass[10pt]{article}
\usepackage[utf8]{inputenc}
\usepackage{cite}
\usepackage{hyperref}
\hypersetup{colorlinks=true,linkcolor=bluecyan,citecolor=orange3,urlcolor=green3,pdfencoding=auto,linktocpage}
\usepackage{cancel}
\usepackage{amsmath,amssymb,amsbsy,amstext,amsthm,simplewick,amsfonts,braket}
\usepackage{braket}
\usepackage{mathrsfs}
\usepackage{graphicx}
\usepackage{wrapfig}
\usepackage{upgreek}
\usepackage{bm} 
\usepackage{framed}
\usepackage{bbm}
\usepackage{textcomp}
\usepackage{adjustbox}
\usepackage{makecell}
\usepackage{tcolorbox}
\usepackage{empheq}
\usepackage[normalem]{ulem}
\usepackage{enumitem}
\usepackage{array}
\usepackage{dsfont}
\usepackage{ulem}
\usepackage{xcolor}
\usepackage{caption}
\usepackage{subcaption}
\usepackage{tocloft}
\usepackage{tikz}
\usepackage[compat=1.1.0]{tikz-feynman}

\usepackage[framemethod=default]{mdframed}

\newmdenv[backgroundcolor=gray!15,%
skipabove=5pt,%
skipbelow=5pt,%
leftmargin=2pt,%
rightmargin=2pt,%
innertopmargin=-6pt,%
innerbottommargin=5pt,%
innerleftmargin=5pt,%
innerrightmargin=5pt,%
splittopskip=0pt,%
splitbottomskip=0pt,%
linewidth=0pt,%
nobreak=true]%
{keyeqn}

\newmdenv[backgroundcolor=gray!15,%
skipabove=5pt,%
skipbelow=5pt,%
leftmargin=2pt,%
rightmargin=2pt,%
innertopmargin=-2pt,%
innerbottommargin=5pt,%
innerleftmargin=5pt,%
innerrightmargin=5pt,%
splittopskip=0pt,%
splitbottomskip=0pt,%
linewidth=0pt,%
nobreak=true]%
{keythrm}

\graphicspath{{Fig/}}

\definecolor{lightgreen}{cmyk}{0.2, 0, 0.2, 0.2}
\definecolor{lightgray}{cmyk}{0.1,0.2,0,0.1}
\definecolor{lightgray2}{cmyk}{0.1,0.1,0,0.1}
\definecolor{bluecyan}{RGB}{0, 100, 200}
\definecolor{blue3}{RGB}{31,119,180}
\definecolor{red3}{RGB}{214,39,40}
\definecolor{orange3}{RGB}{255,127,14}
\definecolor{green3}{RGB}{44,160,44}

\definecolor{red2}{RGB}{255,0,0}
\definecolor{green2}{RGB}{0,170,0}
\definecolor{blue2}{RGB}{0,128,255}
\definecolor{magenta2}{RGB}{191,64,191}
\definecolor{purple2}{RGB}{112,48,160}
\definecolor{orange2}{RGB}{255,192,0}

\newcommand\Ccancel[2][red]{
    \let\OldcancelColor\CancelColor
    \renewcommand\CancelColor{\color{#1}}
    \cancel{#2}
    \renewcommand\CancelColor{\OldcancelColor}
}

\numberwithin{equation}{section}

\allowdisplaybreaks[1]
\begin{document}

\begin{titlepage}
	\setcounter{page}{1} \baselineskip=15.5pt 
	\thispagestyle{empty}

 \begin{center}
		{\fontsize{18}{18}\centering {\bf{Soft Theorems for Boostless Amplitudes}}\;}\\
	\end{center}
 
	\vskip 18pt
	\begin{center}
		\noindent
		{\fontsize{12}{18}\selectfont Zong-Zhe Du\footnote{\tt zongzhe.du@nottingham.ac.uk}$^{,a}$ and David Stefanyszyn\footnote{\tt david.stefanyszyn@nottingham.ac.uk}$^{,a,b}$}
	\end{center}
	
	\begin{center}
		\vskip 8pt
		$a$ \textit{School of Physics and Astronomy,
			University of Nottingham, University Park, \\ Nottingham, NG7 2RD, UK} \\
		$b$ \textit{School of Mathematical Sciences,
			University of Nottingham, University Park, \\ Nottingham, NG7 2RD, UK}  
	\end{center}
	

	\noindent \textbf{Abstract}  We consider effective field theories (EFTs) of scalar fields with broken Lorentz boosts, which arise by taking the decoupling and flat-space limits of the EFT of inflation, and derive constraints that must be satisfied by the corresponding scattering amplitudes if there is an underlying non-linearly realised symmetry. We primarily concentrate on extended shift symmetries which depend on the space-time coordinates, and find that \textit{combinations} of scattering amplitudes obey enhanced Adler zeros. That is, such combinations vanish as one external momentum is taken soft, with the rate at which they vanish dictated by the corresponding symmetry. In our soft theorem derivation, we pay particular care to the energy and momentum-conserving delta functions that arise due to space-time translations, and show that when acted upon by derivatives with respect to spatial momenta, they yield a tower of soft theorems which are ultimately required for closure of the underlying symmetry algebra. All of our soft theorems correspond to constraints that must be satisfied by on-shell amplitudes and, even for symmetries that depend on the time coordinate, our soft theorems only require derivatives to be taken with respect to spatial momenta. We perform a soft bootstrap procedure to find solutions to our soft theorems, and compare these solutions to what we find from an off-shell analysis using the coset construction.

	
\end{titlepage} 


\newpage
\setcounter{page}{2}
{
	\tableofcontents
}


\section{Introduction}

Non-linear realisations and their associated constraints on scattering amplitudes have provided a very useful way of classifying scalar field effective field theories (EFTs) \cite{Cheung:2014dqa}. Off-shell, non-linear symmetries constrain the coupling constants of EFTs by sometimes fixing them to zero, while at other times dictating relations between them. An example of the former is the Galileon symmetry that forbids a $(\partial_\mu \phi)^4$ vertex \cite{Nicolis:2008in,Goon:2012dy}, while an example of the latter is the scalar DBI theory where the couplings of all $(\partial_\mu \phi)^{2n}$ vertices are fixed by the non-linear realisation of the five-dimensional Poincar\'{e} group (see e.g. \cite{deRham:2010eu}).\footnote{The primary difference between these two possibilities is whether the non-linear symmetries of $\phi$ contain field dependence or not: only with field dependence can a symmetry enforce non-trivial relations between coupling constants. Indeed, for the above examples the Galileon theory is characterised by invariance under $\delta \phi = b_{\mu}x^{\mu}$, while the DBI theory is characterised by $\delta \phi = b_{\mu}x^{\mu} + b_{\mu} \phi \partial^{\mu} \phi$ (there is also a constant shift symmetry $\delta \phi = a$ in both cases).} In fact, scalar EFTs with linearly realised four-dimensional Poincar\'{e} symmetries, a constant shift symmetry, and additional symmetries that impose relations between coupling constants are rare, and from this off-shell perspective this can be best understood by classifying the types of algebras that can be realised, as done in \cite{Bogers:2018zeg,Roest:2019oiw}. For scalars there are only two such theories, which are the aforementioned DBI theory and the special Galileon \cite{Hinterbichler:2015pqa}, while for a spin-$1/2$ fermion there is only a single one which is the Volkov-Akulov theory \cite{Roest:2019oiw} (there is a richer structure if we allow for additional fields and supersymmetry \cite{Roest:2019dxy}). 

All of these properties can also be understood on-shell at the level of scattering amplitudes. The constant shift symmetry leads to the Adler zero condition that dictates that amplitudes vanish when one external momentum is sent to zero \cite{Adler:1965ga}. Additional symmetries yield enhanced Adler zeros with the dependence on the coordinates in the symmetry dictating how quickly the amplitude vanishes: a symmetry of the form $\delta \phi = b_{\mu_{1} \ldots \mu_{m}}x^{\mu_{1}} \ldots x^{\mu_{m}} + \ldots$ dictates that $\mathcal{A}_{n} = \mathcal{O}(p^{m+1})$ \cite{Cheung:2014dqa,Cheung:2016drk}. Note that this soft behaviour holds for both field-dependent and field-independent symmetries, with the difference residing in how the soft scaling is realised: in the absence of field dependence each topology contributing to an amplitude satisfies the enhanced Adler zero, while if there is field dependence then cancellations between different topologies are required. EFTs can then be classified based on the number of momenta in the hard amplitude, and the rate at which the soft amplitude vanishes \cite{Cheung:2016drk}. With linearly realised four-dimensional Poincar\'{e} symmetries this story is well-understood. See \cite{Padilla:2016mno,Cheung:2015ota,Cheung:2018oki,Bonifacio:2019rpv,Brauner:2022ymm,CarrilloGonzalez:2019fzc,Bartsch:2022pyi,Goon:2020myi} for other work where Poincar\'{e} invariance is assumed. 

Far less is understood when some of the Poincar\'{e} symmetries are broken, however. One very interesting possibility is where the only broken symmetries are Lorentz boosts with spacetime translations and rotations kept intact thereby allowing us to set-up consistent scattering problems. This set-up is neatly related to de Sitter space and inflationary cosmology. For example, such theories arise by taking the decoupling limit and flat-space limit of the EFT of inflation \cite{Cheung:2007st} (see e.g. \cite{InflationaryAdler} for a recent discussion). More concretely, such boost-breaking scattering amplitudes are contained within cosmological correlators (or more precisely within inflationary wavefunction coefficients) in a particular singular limit. Indeed, (almost) all wavefunction coefficients, from which cosmological correlators can be derived by simple algebraic relations (see e.g. \cite{WFCtoCorrelators1,WFCtoCorrelators2}), are singular when the sum of the magnitudes of the external spatial momenta are taken to zero. It is tempting to associated these magnitudes with energies, but in fact there are no energies or indeed energy conservation in cosmology due to the breaking of time translations, but if the (analytically continued) momenta are tuned such that we realise ``energy conservation", we encounter singularities and these are usually poles if we consider tree-level correlators of the Goldstone boson in the EFT of inflation. On these poles we recover boost-breaking amplitudes, with the relation between wavefunction coefficients and amplitudes given by \cite{Raju:2012zr,Maldacena:2011nz,Pajer:2020wnj}
\begin{align}
\lim_{k_{T} \rightarrow 0} \psi_{n} = (\text{normalisation factors}) \times \frac{\mathcal{A}_{n}}{k_{T}^{p}} \,,
\end{align}
where $\psi_n$ is an inflationary wavefunction coefficient, $\mathcal{A}_{n}$ is a flat-space scattering amplitude, and we write the magnitudes as $k_{a}$ with $k_{T} = \sum_{a=1}^{n} k_{a}$ being the ``total energy". The order of the pole is fixed by the number of derivatives in the theory \cite{BBBB}. If the wavefunction coefficients are really inflationary i.e. they arise from theories with broken de Sitter boosts, the resulting amplitudes are of the boost-breaking type we described above, and which were studied extensively in \cite{Pajer:2020wnj}. If de Sitter boosts are unbroken, then the corresponding amplitudes are Lorentz invariant. This relationship between cosmology and flat-space amplitudes has played an important role in the cosmological bootstrap, see \cite{Baumann:2022jpr,Benincasa:2022omn,Benincasa:2022gtd} for reviews, and \cite{Bonifacio:2022vwa,CosmoBootstrap3,Baumann:2021fxj,MLT} for examples of where scattering amplitudes have been used as input data for such a bootstrap procedure. 

In our quest to understand non-linearly realised symmetries in cosmology and how they affect wavefunction coefficients and cosmological correlators, we are therefore motivated to fully understand boost-breaking amplitudes and how they encode the details of non-linear realisations. A number of interesting works have already been dedicated to understanding non-linear symmetries in the absence of linearly realised Poincar\'{e} symmetries: non-linear realisations in exact de Sitter space have been studied in \cite{Bonifacio:2021mrf,Bonifacio:2019hrj,Bonifacio:2018zex}, algebraic classifications have been performed in \cite{GJS,Mojahed:2022nrn}, the effects of symmetries on wavefunction coefficients in flat-space have been studied in \cite{Bittermann:2022nfh}, non-linear realisations and unequal time correlation functions were studied in \cite{Hui:2022dnm}, while soft theorems for boost-breaking amplitudes were studied in \cite{InflationaryAdler,Cheung:2023qwn}. Fermions with shift symmetries in de Sitter have also been studied in \cite{Bonifacio:2023prb}. So, what are we aiming to add to the discussion? Our primary interest is a toy set-up where the non-linear symmetries are of the Abelian-type i.e. without any field dependence (although our soft theorems will hold more generally). This is a toy set-up since we know that non-linear boosts necessarily come with field dependence due to the non-Abelian nature of the commutators of Lorentz boosts, but as we will see this set-up is rich enough to yield interesting structures. Within this set-up, we derive soft theorems that the corresponding boost-breaking amplitudes must satisfy, and we allow for symmetries that treat space and time separately. We want the solutions to our soft theorems to capture two important properties that are easy to understand off-shell i.e. at the level of the symmetry algebra or at the level of the Lagrangian, but are non-trivial on-shell: 
\begin{itemize}
    \item \textbf{Tower structure}: at the level of the symmetry algebra, the presence of linearly realised space-time translations dictates a tower structure where certain symmetries at some order in the space-time coordinates can only be realised in addition to lower-order symmetries. If we have a symmetry of the schematic form $\delta \pi \sim t^{L}x^{M}$, then closure of the algebra requires all symmetries of the form  $\delta \pi \sim t^{l}x^{m}$ with $l = 0, \ldots, L$ and $m = 0, \ldots, M$. Decreasing the power of $t$ comes from acting with time translations, while decreasing the power of $x$ comes from acting with spatial translations. \textit{We would like our soft theorems to impose this tower structure}. 

    \item \textbf{Invariance of the free theory}: a necessary condition for invariance of some theory under the Abelian-type symmetries we are considering is invariance of the free theory (since operators with different powers of $\pi$ cannot cancel with each other).\footnote{We will use $\pi$ for the fluctuation around the symmetry breaking vev to match with the inflationary literature.} In this paper we are considering a single scalar field so without loss of generality we can take its free theory to be the usual two-derivative Lorentz invariant one since any speed of sound can be fixed to unity by rescaling the spatial coordinates. We therefore have $\mathcal{L}_{2} = \frac{1}{2} \dot{\pi}^2 - \frac{1}{2} (\partial_i \pi)^2$. Demanding invariance of the $\dot{\pi}^2$ part restricts us to consider symmetries that are at most linear in the time coordinate, while invariance of the $ (\partial_i \pi)^2$ part requires the symmetry parameters to be traceless. \textit{We would like our soft theorems to impose that the symmetry parameters are traceless}. Furthermore, we would like the soft theorem to be sensitive to the possibility of traceless symmetry parameters even for contact diagram contributions to scattering amplitudes since even there the amplitude itself is sensitive to the on-shell conditions that we will impose.  
\end{itemize}
As we will see, although we are asking quite simple questions within a toy set-up, we find that the results are quite subtle and lead to interesting structures. We are able to find our desired tower structure by carefully dealing with derivatives acting on the delta functions that impose energy and momentum conservation, and we find that non-trivial solutions to our soft theorems impose traceless conditions on the symmetry parameters, even at the level of four-point contact diagrams. 

One very important difference between the Lorentz-invariant and boost-breaking set-ups is that a constant shift symmetry (which we will assume throughout) within the former set-up does not permit non-trivial cubic vertices i.e. the on-shell three-point amplitude vanishes. This ensures that we don't encounter any poles in the soft limit of $n$-point amplitudes. With broken boosts, however, non-zero on-shell three-point amplitudes are certainly allowed even in the presence of a shift symmetry. A simple example is $\mathcal{A}_{3} = E_{1}E_{2}(E_{1}+E_{2})$ which comes from a $\dot{\pi}^3$ vertex. This causes complications when deriving soft theorems but this problem was tackled in \cite{InflationaryAdler,Cheung:2023qwn} where the consequences of the cubic vertex have to be subtracted using knowledge of the corresponding off-shell vertex. In this paper we will impose a $\pi \rightarrow -\pi$ symmetry such that we don't have the additional complications of cubic vertices. We already encounter a number of subtleties associated with energy and momentum-conserving delta functions and we don't want to muddy the waters by further adding subtleties associated with cubic vertices. In practice, the restriction we will impose is that the symmetry current we will use to derive a soft theorem does not contain terms that are quadratic in $\pi$. This would of course occur if there were cubic vertices but could also arise from field dependence in the symmetry transformation. In this sense our soft theorems also hold for non-Abelian algebras (field-dependent symmetries) as long as there are no quadratic terms in the current i.e. as long as the $\pi \rightarrow - \pi$ symmetry is respected. In an upcoming paper we will drop this assumption \cite{PeterDavid2}. 

We begin our soft theorem derivation in Section \ref{sec:derivation} where we use current conservation and the LSZ reduction formula to analyse the effects of symmetries on on-shell amplitudes. We pay particular care to the ever-present delta functions that impose energy and momentum conservation and find that they are crucial in realising our desired tower structure, but also introduce a neat structure in our soft theorems. Our soft theorems do not contain any ambiguity in how they should be understood when acting on on-shell amplitudes: our soft theorems require one to take derivatives with respect to spatial momenta of combinations of on-shell amplitudes (with energy and momentum conservation imposed) followed by taking a soft limit. The order of first writing amplitudes in terms of a minimal basis, which we detail below, taking derivatives and then taking the soft limit is clear from our derivation. In all cases, even when the symmetries depend on time, our soft theorems only ever require one to take derivatives of on-shell amplitudes with respect to spatial momenta, as opposed to with respect to energies. In this section we also present the general solutions to our soft theorems in terms of soft amplitudes, and show that we recover known Lorentz-invariant results. In Section \ref{sec:bootstrap} we put our solutions to the test by performing a soft bootstrap procedure whereby we construct non-soft amplitudes that yield the desired soft amplitude solutions. We then compare these results with what we would expect from an off-shell Lagrangian analysis with the help of the coset construction with details given in Appendix \ref{app:cosets}. We find agreement between the two methods. We conclude and discuss possible future directions in Section \ref{sec:conclusion}. 

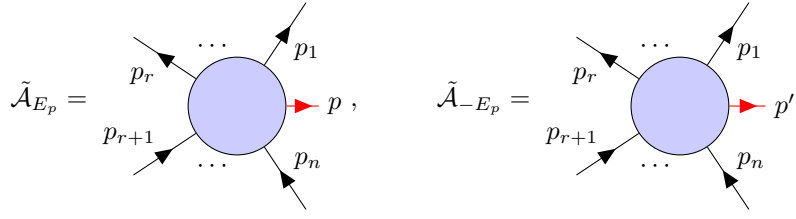
\begin{figure}
    \centering
\[ \tilde{\mathcal{A}}_{E_{p}} = 
\begin{tikzpicture}[baseline=(m.base)]
    \begin{feynman}
      \vertex[draw,circle,fill=blue!20,minimum size=1.3cm] (m) at ( 0, 0) {};
      \vertex (a) at (-1.5,-1) {};
      \vertex (b) at ( 1,-1.5) {};
      \vertex (c) at (-1.5, 1) {};
      \vertex (d) at ( 1, 1.5) {};
      \vertex (e) at ( -0.3, 0.8) {$\cdots$};
      \vertex (f) at ( -0.3, -0.8) {$\cdots$};
      \vertex (g) at ( 1.3, 0) {$p$};
      \diagram* {
        (a) -- [fermion,edge label=$p_{r+1} $] (m) -- [fermion,edge label=$p_{r}$] (c),
        (b) -- [fermion,edge label'=$p_{n}$] (m) -- [fermion,edge label'=$p_{1} $] (d),
        (m) -- [fermion, red] (g)
      };
    \end{feynman}
  \end{tikzpicture} 
,\;\;\;\qquad \tilde{\mathcal{A}}_{-E_{p}} = 
  \begin{tikzpicture}[baseline=(m.base)]
    \begin{feynman}
      \vertex[draw,circle,fill=blue!20,minimum size=1.3cm] (m) at ( 0, 0) {};
      \vertex (a) at (-1.5,-1) {};
      \vertex (b) at ( 1,-1.5) {};
      \vertex (c) at (-1.5, 1) {};
      \vertex (d) at ( 1, 1.5) {};
      \vertex (e) at ( -0.3, 0.8) {$\cdots$};
      \vertex (f) at ( -0.3, -0.8) {$\cdots$};
      \vertex (g) at ( 1.4, 0) {$p'$};
      \diagram* {
        (a) -- [fermion,edge label=$p_{r+1} $] (m) -- [fermion,edge label=$p_{r}$] (c),
        (b) -- [fermion,edge label'=$p_{n}$] (m) -- [fermion,edge label'=$p_{1} $] (d),
        (m) -- [fermion, red] (g)
      };
    \end{feynman}
  \end{tikzpicture} 
\]
    \caption{On the LHS we have the graphical representation of an $(n+1)$-point amplitude $\tilde{\mathcal{A}}_{E_{p}}$ with $n$ hard momenta denoted by $p_{1}$ to $p_{n}$ and one soft momentum denoted by $p = (E_{p}, \Vec{p})$. The tildes indicate that energy and momentum conservation have been imposed. The figure on the RHS is a graphical representation of $\tilde{\mathcal{A}}_{-E_{p}}$ which is the same amplitude but with the energy of the soft leg flipped i.e. $p'=(-E_{p},\Vec{p})$. All hard momenta are identical to those in $\tilde{\mathcal{A}}_{E_{p}}$.} 
    \label{fig:GraphicRep}
\end{figure}

\paragraph{Summary of results} Before diving into our soft theorem derivation, let us summarise our main results here for the benefit of the reader. We have two sets of soft theorems: one for symmetries that only depend on the spatial coordinates i.e. $\delta \pi = b_{i_{1} \ldots i_{n}}x^{i_{1}} \ldots x^{i_{n}}$ which take the form
\begin{keyeqn}
\begin{align}
b_{i_{1} \ldots i_{n}}  \lim_{\vec{p} \rightarrow 0} \biggl\{\partial_{p_{i_1}}\dots\partial_{p_{i_k}}\left[\tilde{\mathcal{A}}_{E_{p}}+\tilde{\mathcal{A}}_{-E_{p}} \right] \biggr\} & =0\,,k=0,1,2,\ldots,n\,,\\
    b_{i_{1} \ldots i_{n}}   \lim_{\vec{p} \rightarrow{}0}  \biggl\{ \partial_{p_{i_1}}\dots\partial_{p_{i_k}} \left[\left (\tilde{\mathcal{A}}_{E_{p}} - \tilde{\mathcal{A}}_{-E_{p}}\right) E_{p} \right] \biggr\}&=0\,,k=0,1,2,\ldots,n\,, \label{Summaryeqn1}
\end{align}
\end{keyeqn}
and another set for symmetries that depend linearly on the time coordinate i.e. $\delta \pi = c_{i_{1} \ldots i_{n}} t x^{i_{1}} \ldots x^{i_{n}}$ which take the form
\begin{keyeqn}
\begin{align}
c_{i_{1} \ldots i_{n}} \lim_{\vec{p} \rightarrow 0} \biggl\{\partial_{p_{i_1}}\dots\partial_{p_{i_k}}\left[ \tilde{\mathcal{A}}_{E_{p}}+ \tilde{\mathcal{A}}_{-E_{p}} \right] \biggr\} & = 0\,, k=0,1,2,\ldots,n \,, \\ 
c_{i_{1} \ldots i_{n}} \lim_{\vec{p} \rightarrow 0} \biggl\{\partial_{p_{i_1}}\dots\partial_{p_{i_k}}\left[\frac{\tilde{\mathcal{A}}_{E_{p}}-\tilde{\mathcal{A}}_{-E_{p}}}{E_{p}} \right] \biggr\} & =0\,, k=0,1,2,\ldots,n \,. \label{Summaryeqn2}
\end{align}
\end{keyeqn}
In these expressions a tilde represents that energy and momentum conservation have been imposed such that $\tilde{A}_{E_{p}}$ is the on-shell amplitude with all conditions imposed and without the delta functions, while $\tilde{A}_{-E_{p}}$ is the same function but with $E_{p} \rightarrow -E_{p}$, where $E_{p}$ is the energy associated with the momentum that we take soft. These amplitudes are shown in Figure \ref{fig:GraphicRep}. We will usually take this soft momentum to be $p_{1}^{\mu}$ so in the above expressions we have $p^{\mu} \equiv p^{\mu}_{1}$. We note that for a given $n$, any soft amplitudes that solve \eqref{Summaryeqn2}, automatically solve \eqref{Summaryeqn1} with the same $n$, which is precisely what is required by the tower structure given that the corresponding symmetries are related by a time translation.

\paragraph{Minimal basis} When computing scattering amplitudes we will work in the minimal basis where all on-shell conditions and energy and momentum conservation have been taken into account. Here we outline this minimal basis and refer the reader to  \cite{Cheung:2023qwn} for more details. We take all particles as incoming for simplicity. For an $n$-point amplitude, the on-shell external four-momenta are
\begin{align}
    p^{\mu}_{a} = (E_{a}, \Vec{p}_{a}) \;\; \text{for}\;\; 1 \leq a \leq n \,,
\end{align}
 where $E_{a}\equiv |\Vec{p}_{a}|$ such that $p^2_{a} = 0$. Since we are interested in scattering processes that are constrained by $SO(3)$ symmetry, there are in principle $\frac{n(n+1)}{2}$ invariant building blocks:
\begin{align}
    E_{a}\;\; &\text{for} \;\; 1 \leq a \leq n \,, \\  
    \Vec{p}_{b} \cdot \Vec{p}_{c} \;\;& \text{for} \;\; 1\leq b< c \leq n \,.
\end{align}
To impose energy and momentum conservation, we eliminate $\Vec{p}_{n}$ and $E_{n} $ by
\begin{align}
    \vec{p}_{n} &= - \sum_{a=1}^{n-1} \vec{p}_{a}\,, \\
    E_{n} &=- \sum_{a=1}^{n-1} E_{a} \,.
\end{align}
In terms of the invariant building blocks, these constraints can be used to eliminate $E_{n}$, and $\Vec{p}_{a}\cdot\Vec{p}_{n}$ with $a \neq n $. Moreover, the fact that $p_{n}$ is on-shell allows us to eliminate one further building block using 
\begin{align}
\left( \sum_{a=1}^{n-1} E_{a} \right)^2 = \left(\sum_{a=1}^{n-1} \vec{p}_{a} \right)^2\,.
\end{align}
We use this constraint to eliminate $\Vec{p}_{n-2}\cdot\Vec{p}_{n-1}$. We have now imposed all conditions and are left with the following invariant building blocks which form the minimal basis:
\begin{keyeqn}
\begin{align}
    E_{a} \;\; &\text{for} \;\;1 \leq a \leq n-1 \,,\\
    \vec{p}_{b} \cdot \Vec{p}_{c} \;\; &\text{for} \;\;1 \leq b < c \leq n-2 \;\;\text{and}\;\; 1 \leq b \leq n-3,\;c = n-1 \,.
\end{align}
\end{keyeqn}
There are a total of $\frac{n(n-1)}{2} - 1$ such building blocks. As an example, for $n=3$ the minimal basis is simply
\begin{align}
  \{ E_{1},\; E_{2} \} \,,
\end{align}
while for $n=4$ we have
\begin{align}
  \{ E_{1},\; E_{2},\; E_{3},\; \Vec{p}_{1} \cdot \Vec{p}_{2},\; \Vec{p}_{1} \cdot \Vec{p}_{3} \} \,.
\end{align}
In this case, $\vec{p}_{2} \cdot \vec{p}_{3}$ has been eliminated using the on-shell condition for $\vec{p}_{4}$ which fixes $\Vec{p}_{2}\cdot \Vec{p}_{3} = E_{2} E_{3} + E_{1} E_{2} +E_{1} E_{3} - \vec{p}_{1}\cdot \vec{p}_{2} - \vec{p}_{1}\cdot \vec{p}_{3}$.


\section{Soft theorems and solutions} \label{sec:derivation}
We begin our derivation with the Ward-Takahashi identity relating correlation functions arising from a Lagrangian with a global symmetry $\pi(x) \rightarrow \pi(x) + \epsilon \delta \pi(x)$, and Noether current $J^{\mu}(x)$: 
\begin{equation} 
    \partial_\mu\braket{0|\mathcal{T}\{J^\mu(x)\pi(x_1) \ldots \pi(x_n)\}|0} = -  i\sum_{a=1}^{n}\delta^{(4)}(x-x_a)\braket{0|\mathcal{T}\{\pi(x_1) \ldots\delta\pi(x_a)\ldots\pi(x_n)\}|0},
    \label{WTidentity}
\end{equation}
where $\partial_\mu$ acts on coordinates $x^{\mu}$, $\mathcal{T}$ represents the usual time-ordering, and we have expanded to linear order in $\epsilon$. Our aim is to derive conditions that $\textit{on-shell}$ amplitudes satisfy as a consequence of the presence of such a global symmetry, so we use the LSZ reduction formula on \eqref{WTidentity}. As we explained in the introduction, our aim in this work is to derive the soft theorems associated with field-independent symmetry transformations (and therefore with Abelian structures for the commutators between broken generators) for which $\delta \pi(x)$ is a function of the coordinates only. In more familiar Lorentz-invariant cases, this restriction would capture Galileon field theories, but not DBI ones (see e.g \cite{Bogers:2018zeg,Roest:2019oiw}). With this restriction, the RHS of \eqref{WTidentity} does not generate any poles in the variables associated with the Fourier transform of the coordinates $x_{a}$, and therefore there will be no contribution from the RHS once we apply the LSZ reduction. Indeed, the LSZ reduction formula takes an $n$-point correlator, Fourier transforms in each coordinate and puts all momenta on-shell. The coefficient of the most singular part of the result is an $S$-matrix element with $r$ particles in the $\textit{out}$ state and $n-r$ particles in the $\textit{in}$ state.\footnote{If we have field dependence in $\delta \pi$ then one can still argue that the RHS gives a vanishing contribution since the pole associated with the $\delta \pi$ insertion does not correspond to a single external momentum going on-shell. The vanishing of the RHS requires one to take the on-shell limit prior to the soft limit, see e.g. \cite{Green:2022slj}.} We therefore concentrate of the LHS of \eqref{WTidentity} from now on, and will express the LSZ operation as
\begin{align}
\prod_{a=1}^r\prod_{b=r+1}^n\text{LSZ}_{a+}\text{LSZ}_{b-}&\left[f(\pi(x_{a}), \pi(x_{b})) \right] \nonumber \\ &\equiv\prod_{a=1}^r \prod_{b=r+1}^n\lim_{p_a^0\rightarrow E_{a}} \lim_{p_b^0\rightarrow E_{b}} \int_{x_a} \int_{x_b} \,e^{ip_a \cdot x_a} \,e^{-ip_b \cdot x_b} p_a^2 p_b^2 \, f(\pi(x_{a}), \pi(x_{b}))\,,
\end{align}
where $p^\mu_a=(p^0_a, \vec{p}_{a})$ denotes the four-momentum of particle $a$, $p_a^2 =  \eta_{\mu\nu}p_a^\mu p_a^\nu$, $p_a \cdot x_a= \eta_{\mu\nu}p_a^\mu x_{a}^\nu$, $\int_{x_a} \equiv \int d^4x_a$ and $E_{a}\equiv |\Vec{p}_{a}|$. To analyse this operation on the LHS of \eqref{WTidentity}, we split the time integrals into three regions:
\begin{align}
\text{Far-past Region}:& ~~ t<T^-\\
\text{Intermediate Region}:& ~~ T^- \leq t \leq T^+\\
\text{Far-future Region}:& ~~ t>T^+
\end{align}
where $T^-$ and $T^+$ are respectively the early and late time slices which before and after the particles are asymptotically free. Singularities associated with one-particle states can only arise from the regions where $t<T^-$ and $t>T^+$, and so for all time integrals we can ignore the intermediate regions where $T^-\leq t \leq T^+$. Since we take the energies to be positive such that the on-shell conditions are enforced by $p^0\rightarrow E_{p}$ rather than $p^0\rightarrow - E_{p}$, poles associated with particles in the out state only come from the $t>T^+$ region, while poles associated with particles in the in state only come from the $t<T^-$ region. We can therefore work with the following more compact form of the LSZ operation:
\begin{align}
\prod_{a=1}^r\prod_{b=r+1}^n & \text{LSZ}_{a+}\text{LSZ}_{b-}\left[f(\pi(x_{a}), \pi(x_{b})) \right] \nonumber \\ &\equiv\prod_{a=1}^r \prod_{b=r+1}^n\lim_{p_a^0\rightarrow E_{a}} \lim_{p_b^0 \rightarrow E_{b}} \int_{t_a > T^+} \int_{\vec{x}_{a}} \int_{t_b < T^-}  \int_{\vec{x}_{b}} \,e^{ip_a\cdot x_a} \,e^{-ip_b\cdot x_b} p_a^2 p_b^2 \, f(\pi(x_{a}), \pi(x_{b}))\,.
\end{align}
The Ward-Takahashi identity then becomes
\begin{keyeqn}
    \begin{align} \label{WTidentity2}
    \prod_{a=1}^r\prod_{b=r+1}^n\text{LSZ}_{a+} \text{LSZ}_{b-}\braket{0|\pi(x_1)\pi(x_2)\ldots \pi(x_r)\partial_\mu J^\mu(x)\pi(x_{r+1})\pi(x_{r+2})\dots \pi(x_n)|0}=0\,.
    \end{align}
    \end{keyeqn}
Note that we have not Fourier transformed the current contribution. To proceed we need to specify a form for the current which always satisfies 
\begin{align}
\partial_{\mu}J^{\mu} (x) = \delta \pi \Box \pi + \mathcal{O}(\pi^3)\,,
\end{align}
thanks to our freedom to take the kinetic term to be Lorentz invariant, and since we impose a $\pi \rightarrow - \pi$ symmetry. 

\paragraph{Constant shift symmetry ($\delta \pi = a$)} Initially, consider the simplest case of a constant shift symmetry $\delta \pi = a$ in addition to $\pi \rightarrow - \pi$. We then have
\begin{align} \label{AdlerZeroCurrent}
\partial_{\mu} J^{\mu} (x) = a \Box \pi + \mathcal{O}(\pi^3). 
\end{align}
If we plug this expression into \eqref{WTidentity2}, Fourier transform, and take the momentum associated with $x$ to zero, we have  
 \begin{align} \label{AdlerZeroCurrent1}
\lim_{q \rightarrow 0 } \left[\int_{x} q^2 e^{i q \cdot x} \prod_{a=1}^r\prod_{b=r+1}^n\text{LSZ}_{a+} \text{LSZ}_{b-}\braket{0|\pi(x_1)\ldots \pi(x_r) \pi(x) \pi(x_{r+1})\dots \pi(x_n)|0} \right]=0\,.
    \end{align}
Taking $q_{\mu} \rightarrow 0$ has allowed us to drop all non-linear terms in \eqref{AdlerZeroCurrent1} since they are sub-dominant in this soft limit \cite{Green:2022slj}. Note that at this stage $q_{\mu}$ is not on-shell; this expression is more general. However, we can go on-shell without spoiling this condition in which case we have
\begin{align}
\lim_{q \rightarrow 0 }\left[ \text{LSZ}_{x+}   \prod_{a=1}^r\prod_{b=r+1}^n\text{LSZ}_{a+} \text{LSZ}_{b-}\braket{0|\pi(x_1)\ldots \pi(x_r) \pi(x) \pi(x_{r+1})\dots \pi(x_n)|0} \right]=0\,,
    \end{align}
where to go on-shell we have set $q^{0} \rightarrow E_{q}$. Since the LSZ reduction has now been applied to all $\pi$'s in the correlator, we can write this as a constraint on the $(n+1)$-point amplitude:
\begin{align} \label{AdlerZero1}
\lim_{q \rightarrow 0 } \left[ \mathcal{A}_{n+1}(\vec{q}, \{\vec{p}_{a}\},\{\vec{p}_{b}\} ) \delta\left(E_{q} + \sum_{a=1}^{r} E_{a} - \sum_{b=r+1}^{n} E_{b} \right) \delta^{(3)}\left(\vec{q} + \sum_{a=1}^{r} \vec{p}_{a} - \sum_{b=r+1}^{n} \vec{p}_{b} \right) \right]=0\,,
\end{align}
where $\{ \vec{p}_{b} \}$ and $\{ \vec{p}_{a} \}$ collectively denote the momenta of in and out states respectively. This is the usual Adler zero condition that states that on-shell amplitudes, with energy and momentum conservation imposed, vanish in the soft limit if there is an underlying constant shift symmetry. Note that this derivation did not assume Lorentz invariance of the interactions so the Adler zero holds for boost-breaking amplitudes too.

To arrive at a condition on the fully on-shell amplitude, we put $q_{\mu}$ on-shell by hand since the condition we derived was valid for all $q_{\mu}$. If the Ward-Takahashi identity involves derivatives, however, which will be the case when the symmetry transformation has explicit dependence on the coordinates, we cannot simply put $q_{\mu}$ on-shell by hand since ``going on-shell" does not commute with taking the derivative. Let us therefore see how we can still extract the Adler zero condition without explicitly going on-shell. As we did for the LSZ reduction, we again split the time integral in $\int_{x}$ into three regions:
\begin{equation}
\int_x e^{iq\cdot x} q^2 = \int_{\vec{x}} \left[ \int_{t>T^+} e^{iq\cdot x} q^2 + \int_{T^- \leq t \leq T^+} e^{iq\cdot x} q^2+ \int_{t<T^-} e^{iq\cdot x} q^2 \right], 
\end{equation}
such that \eqref{AdlerZeroCurrent1} becomes
\begin{align}
\lim_{q\rightarrow 0} \left[ \int_{\vec{x}} \left(G^{+} + G^{0} + G^{-} \right) \right] = 0\,,
\end{align}
where 
\begin{align}
        G^+ &= \int_{t>T^+} q^2 e^{iq\cdot x}  \prod_{a=1}^r \prod_{b=r+1}^n\text{LSZ}_{a+} \text{LSZ}_{b-}\braket{0|\pi(x_1)\ldots \pi(x_r)\pi(x)\pi(x_{r+1})\dots \pi(x_n)|0}\,,\\
        G^0 &= \int_{T^- \leq t \leq T^+} q^2  e^{iq\cdot x} \prod_{a=1}^r \prod_{b=r+1}^n\text{LSZ}_{a+} \text{LSZ}_{b-}\braket{0|\pi(x_1)\ldots \pi(x_r)\pi(x)\pi(x_{r+1})\dots \pi(x_n)|0}\,,\\
        G^- &= \int_{t<T^-} q^2 e^{iq\cdot x}  \prod_{a=1}^r \prod_{b=r+1}^n\text{LSZ}_{a+} \text{LSZ}_{b-}\braket{0|\pi(x_1)\ldots \pi(x_r)\pi(x)\pi(x_{r+1})\dots \pi(x_n)|0}\,,
\end{align}
and we have adopted the notation of \cite{Gillioz:2020mdd}. The computation of each mirrors that of the LSZ reduction procedure: we insert a complete set of states and perform the various momentum and coordinate integrals. Note that we insert on-shell states even within the intermediate region. For the far-past and far-future regions, all wave packets are well separated, hence each particle can safely be put on-shell. For the intermediate region, the only contribution comes from the soft mode since contributions from hard modes are projected out by multiplying by $p_a^2$, sending $p_a^2 \rightarrow 0$, and noting that there are no singular contributions that can cancel and yield a finite result. For this reason we can take $T^{+} \rightarrow T^{-} = T$ without affecting the result. The contributions from $G^{+}$ and $G^{-}$ contain poles as $q^{0} \rightarrow \pm E_{q}$ which are cancelled by the overall factor of $q^2$ in each term. We then have\footnote{In practice we inserted the one-particle completeness relation, however in principle multi-particle states should also be considered. We drop multi-particle states since they are formally separated, namely their contributions to scattering process are completely different (the delta function structures are different, regardless of the soft momentum, for example).} 
\begin{align} \label{SoftDerivation1}
\lim_{q\rightarrow 0}\left[\frac{q^0+E_{q}}{2E_{q}}e^{i(q^0-E_{q})T}\braket{\alpha,q|\beta} + \frac{-q^0+E_{q}}{2E_{q}}e^{i(q^0+E_{q})T}\braket{\alpha,q'|\beta} \right] = 0\,,
\end{align}
where $\braket{\alpha,q|\beta}= \braket{\vec{p}_1\vec{p}_2\dots\vec{p}_r\vec{q}|\vec{p}_{r+1}\vec{p}_{r+2}\dots\vec{p}_{n}}$ with all $\ket{\vec{p}_c}$ and $\ket{\vec{q}}$ being on-shell states whose energies are $E_{c}$ and $E_{q}$ respectively. $\braket{\alpha,q'|\beta}$ is related to $\braket{\alpha,q|\beta}$ by flipping the sign of $E_{q}$, with the former coming from $G^{-}$ and the latter from $G^{+}$. As an example of this energy flipping, consider the quartic interaction $\frac{g}{4!}\dot{\pi}^4$. We have (taking all particles as incoming)
\begin{align} \label{AmplitudeExample1}
\tilde{\mathcal{A}}^{\dot{\pi}^4}_{E_{1}}  = - gE_{1}E_{2}E_{3}(E_{1}+E_{2}+E_{3})\,,
\end{align}
and therefore
\begin{align} \label{AmplitudeExample2}
\tilde{\mathcal{A}}^{\dot{\pi}^4}_{-E_{1}}  = g E_{1}E_{2}E_{3}(-E_{1}+E_{2}+E_{3})\,.
\end{align} 
These two expressions only differ at linear order in $E_{1}$ (and of course more generally such expressions will only differ at odd orders in $E_{1}$). Here we have introduced the notation that $\tilde{\mathcal{A}}$ is the amplitude with energy and momentum conservation imposed and with the corresponding delta functions stripped off. More generally, we compute $\tilde{\mathcal{A}}_{-E_{1}}$ by going to the minimal basis outlined in the introduction, followed by flipping the sign of the indicated energy once all on-shell conditions and energy and momentum conservation have been applied.  

Again the delta functions that impose energy and momentum conservation are going to play an important role in what follows so let's write them out explicitly. From \eqref{SoftDerivation1} we have  
\begin{align} \label{WTIntermediate}
\lim_{q\rightarrow 0} \biggl\{\left[\frac{q^0+E_{q}}{2E_{q}}e^{i(q^0-E_{q})T}\mathcal{A}_{E_{q}} \delta(E_{q} + p^{0})  + \frac{-q^0+E_{q}}{2E_{q}}e^{i(q^0+E_{q})T}\mathcal{A}_{-E_{q}} \delta(-E_{q} + p^{0})  \right] \delta^{(3)}(\vec{q} + \vec{p}) \biggr\} = 0\,,
\end{align}
where we have defined a new four-vector $p_{\mu}$ with components
\begin{align}
p^{0} &= \sum_{a=1}^{r} E_{a}  - \sum_{b=r+1}^{n} E_{b} \,, \\ 
\vec{p} &= \sum_{a=1}^{r}  \vec{p}_{a}  - \sum_{b=r+1}^{n}  \vec{p}_{b}  \,.
\end{align}
We have suppressed the arguments of $\mathcal{A}$ which are simply the on-shell kinematics, but have used the energy subscript to indicate if the energy associated with the momentum which we will ultimately take soft should be flipped or not. We can now use the presence of the overall delta function $\delta^{(3)}(\vec{q} + \vec{p})$ to eliminate the $\vec{q}$ dependence in the square brackets followed by taking the soft limit. This yields 
\begin{align}
\left[\tilde{\mathcal{A}}_{E_{p}} \delta(E_{p}+p^{0}) + \tilde{\mathcal{A}}_{-E_{p}} \delta(-E_{p}+p^{0})   \right]e^{i p^{0} T} \delta^{(3)}(\vec{p}) = 0\,.
\end{align}
We have added the tildes even though we still explicitly have the energy-conserving delta functions to emphasise that momentum conservation has been taken care of. We can now treat $\vec{p}$ as an independent variable by e.g. eliminating $\vec{p}_{n}$ such that the $(n+1)$-point amplitude with momentum conservation imposed is a function of the following three-momenta: $\vec{p}, \vec{p}_{1}, \ldots \vec{p}_{n-1}$.\footnote{Bose symmetry ensures that this change of variables is consistent and yields a soft theorem that acts on the same amplitude that appears in all previous expressions.} The delta function is then telling us to take the soft limit thereby resulting in
\begin{keyeqn}
\begin{align} \label{AdlerZero2}
\lim_{\vec{p} \rightarrow 0} \biggl\{\tilde{\mathcal{A}}_{E_{p}} + \tilde{\mathcal{A}}_{-E_{p}} \biggr\} = 0\,.
\end{align}
\end{keyeqn}
This recovers the Adler zero condition we derived above: any amplitude which is zeroth order in the soft momentum is ruled out by both \eqref{AdlerZero1} and \eqref{AdlerZero2}. Note that we are taking the spatial momentum to zero, but since this acts on the on-shell amplitude, the energy is also taken to zero. 

\paragraph{Linear in $x^i$ symmetry ($\delta \pi = b_{i} x^{i}$)}
Let us now turn our attention back to symmetries that depend on the coordinates, and begin our discussion with the $\delta \pi = b_{i} x^{i}$ symmetry where
\begin{align}
\partial_{\mu} J^{\mu} (x) = b_{i}x^{i} \Box \pi + \mathcal{O}(\pi^3)\,.
 \end{align}
 A number of the steps we went through above are very similar here so let's jump to the analog of \eqref{WTIntermediate} which takes the form:
 \begin{align}
  \lim_{q\rightarrow 0} \partial_{q_i} \biggl\{ \left[\frac{q^0+E_{q}}{2E_{q}}e^{i(q^0-E_{q})T}\mathcal{A}_{E_{q}}\delta(E_{q}+p^0) + \frac{-q^0+E_{q}}{2E_{q}}e^{i(q^0+E_{q})T}\mathcal{A}_{-E_{q}}\delta(-E_{q}+p^0) \right] \delta^{(3)}(\vec{q}+\vec{p}) \biggr\} = 0\,,
 \end{align}
where the derivative acts on the full bracket, \textit{including} the delta functions. Using the delta function for spatial momentum conservation we can then write
 \begin{align}
  \lim_{q\rightarrow 0} \biggl\{ \left[\frac{q^0+E_{p}}{2E_{p}}e^{i(q^0-E_{p})T}\mathcal{A}_{E_{p}}\delta(E_{p}+p^0) + \frac{-q^0+E_{p}}{2E_{p}}e^{i(q^0+E_{p})T}\mathcal{A}_{-E_{p}}\delta(-E_{p}+p^0) \right] \partial_{p_i}  \delta^{(3)}(\vec{q}+\vec{p}) \biggr\} = 0\,,
 \end{align}
followed by taking the soft limit to yield 
 \begin{align} \label{IntermediateStep}
\left[\tilde{\mathcal{A}}_{E_{p}}\delta(E_{p}+p^0) + \tilde{\mathcal{A}}_{-E_{p}}\delta(-E_{p}+p^0) \right]e^{i p^{0} T} \partial_{p_i}  \delta^{(3)}(\vec{p}) = 0\,.
 \end{align}
 Again, we have included the tildes to indicate that momentum conservation has been imposed. This structure is now reminiscent of what one finds in the Lorentz-invariant case, see e.g. \cite{Cheung:2016drk} (in the sense that we have an amplitude multiplied a derivative acting on a delta function). Using properties of the delta function, we then have two conditions given by
\begin{align}
\lim_{\vec{p}\rightarrow{}0}\biggl\{ \tilde{\mathcal{A}}_{E_{p}} +\tilde{\mathcal{A}}_{-E_{p}}\biggr\}  &= 0 \,, \label{BSymmetryLeading} \\
  \lim_{\vec{p}\rightarrow{}0}  \biggl\{ \partial_{p_i} \left[\tilde{\mathcal{A}}_{E_{p}} e^{i p^{0} T} \delta(E_{p}+p^{0})  + \tilde{\mathcal{A}}_{-E_{p}} e^{i p^{0} T} \delta(-E_{p}+p^{0}) \right]  \biggr\} &=0 \label{BSymmetryEnhanced}\,.
\end{align}
The first of these is simply the Adler zero condition we found before, c.f. \eqref{AdlerZero2}. It is comforting that we found this condition since the presence of a $\delta \pi = b_{i}x^{i}$ symmetry requires the presence of a $\delta \pi = a$ symmetry for the symmetry algebra to close. The relevant commutator is 
\begin{align}
[P_{i}, B_{j}] = \delta_{ij} A\,,
\end{align}
where $B_{i}$ and $A$ generate the above symmetries with parameters $b_{i}$ and $a$. We see that it is the commutator with spatial translations that is important to see the necessity of the constant shift symmetry, and in our soft theorem derivation we see that it is the spatial derivative acting on the delta function of spatial momentum conservation, that itself is a consequence of invariance under spatial translations, that gives rise to multiple constraints with one of these being the Adler zero. We therefore see a nice connection between the algebra and the soft theorem, and the role of spatial translations. A complete soft theorem should indeed capture this \textit{tower structure}. 

Let's now study the other condition we found which is given by \eqref{BSymmetryEnhanced}. As before, we treat the independent three-momenta as $\vec{p}, \vec{p}_{1}, \ldots \vec{p}_{n-1}$ which means that the derivative with respect to $\vec{p}$ acts on both $E_{p}$ and $p^0$. The fact that it acts on $E_{p}$ is obvious since $E_{p} = | \vec{p} | $, while the fact that it acts on $p^{0}$ follows from
\begin{align}
p^0 = \sum_{a=1}^{r} E_{a} - \sum_{b=r+1}^{n-1} E_{b} - \mid \sum_{a=1}^{r} \vec{p}_{a} - \sum_{b=r+1}^{n-1} \vec{p}_{b} - \vec{p} \mid \,.
\end{align}
To proceed, we Taylor expand the delta functions of energy conservation around $\vec{p} = 0$:
\begin{align}
    \delta(\pm E_{p} + p^0) = \delta(\bar{p}^{0}) \pm E_{p} \delta'(\bar{p}^{0}) +\mathcal{O}(\vec{p}\cdot \vec{p}_c) + \mathcal{O}(E_{p}^2) \,,
\end{align}
where 
\begin{align}
\bar{p}^0 = \sum_{a=1}^{r} E_{a} - \sum_{b=r+1}^{n-1} E_{b} - \mid \sum_{a=1}^{r} \vec{p}_{a} - \sum_{b=r+1}^{n-1} \vec{p}_{b}  \mid \,,
\end{align}
and $\vec{p}\cdot \vec{p}_c$ denotes a general dot product between $\vec{p}$ and any of the other momenta. For now we concentrate on the first two terms in these Taylor expansions, then discuss the corrections. In this case \eqref{BSymmetryEnhanced} becomes
\begin{align}
&\lim_{\vec{p}\rightarrow{}0}  \biggl\{  \partial_{p_i} \left [\tilde{\mathcal{A}}_{E_{p}} + \tilde{\mathcal{A}}_{-E_{p}}\right]  \biggr\} \delta(\bar{p}^{0}) e^{i \bar{p}^{0} T} \nonumber \\ 
+ & \lim_{\vec{p}\rightarrow{}0}  \biggl\{ \tilde{\mathcal{A}}_{E_{p}}  + \tilde{\mathcal{A}}_{-E_{p}} \biggr\} \delta(\bar{p}^{0}) \lim_{\vec{p}\rightarrow{}0} \partial_{p_{i}} e^{i p^{0} T} \nonumber \\
+ &  \lim_{\vec{p}\rightarrow{}0}  \biggl\{ \partial_{p_i} \left[ \left(\tilde{\mathcal{A}}_{E_{p}} - \tilde{\mathcal{A}}_{-E_{p}} \right) E_{p} \right]   \biggr\} \delta'(\bar{p}^{0}) e^{i \bar{p}^{0} T}= 0\,.
\end{align}
The middle line vanishes by the leading Adler zero c.f.\eqref{BSymmetryLeading}, and so we are left with two conditions:
\begin{keyeqn}
\begin{align}
 \lim_{\vec{p}\rightarrow{}0}  \biggl\{ \partial_{p_i} \left [\tilde{\mathcal{A}}_{E_{p}} + \tilde{\mathcal{A}}_{-E_{p}}\right]  \biggr\}&=0\,, \label{LinearInx1}\\
 \lim_{\vec{p}\rightarrow{}0}  \biggl\{ \partial_{p_i} \left[ \left(\tilde{\mathcal{A}}_{E_{p}} - \tilde{\mathcal{A}}_{-E_{p}} \right) E_{p} \right]   \biggr\} &=0\, . \label{LinearInx2}
\end{align}
\end{keyeqn}
The higher-order contributions from expanding the delta functions are of the form
\begin{align}
    \lim_{\vec{p}\rightarrow{}0}  \biggl\{ \partial_{p_i} \left[ \left (\tilde{\mathcal{A}}_{E_{p}} + \tilde{\mathcal{A}}_{-E_{p}}\right) E_{p}^{2n} (\vec{p} \cdot \vec{p}_{c})^m \right] \biggr\} \times \text{delta structure}  & \,,\\
     \lim_{\vec{p}\rightarrow{}0}  \biggl\{ \partial_{p_i} \left[\left (\tilde{\mathcal{A}}_{E_{p}} - \tilde{\mathcal{A}}_{-E_{p}}\right) E_{p}^{2n+1} (\vec{p} \cdot \vec{p}_{c})^m \right] \biggr\} \times \text{delta structure}&\,,
\end{align}
which vanish thanks to \eqref{LinearInx1} and \eqref{LinearInx2}, and the Adler zero, and so these two conditions (along with the Adler zero) are necessary and sufficient to satisfy \eqref{BSymmetryEnhanced}.

\paragraph{General purely-spatial symmetry ($\delta \pi  = b_{i_{1} \ldots i_{n}} x^{i_{1}} \ldots x^{i_{n}}$)} The derivation for the more general symmetry that depends on the spatial coordinates only is very similar to what we just went through above. We arrive at \eqref{IntermediateStep} but now with $n$ derivatives with respect to the spatial momenta acting on the momentum-conserving delta function. Being careful with these derivatives and the energy-conserving delta functions yields a system of constraints of the form:
\begin{keyeqn}
\begin{align}
   b_{i_{1} \ldots i_{n}}     \lim_{\vec{p}\rightarrow 0} \biggl\{\partial_{p_{i_1}}\dots\partial_{p_{i_k}}\left[\tilde{\mathcal{A}}_{E_{p}}+\tilde{\mathcal{A}}_{-E_{p}} \right] \biggr\} & =0\,,k=0,1,2,\ldots,n\,,\label{Leading}\\
    b_{i_{1} \ldots i_{n}}   \lim_{\vec{p}\rightarrow{}0}  \biggl\{ \partial_{p_{i_1}}\dots\partial_{p_{i_k}} \left[\left (\tilde{\mathcal{A}}_{E_{p}} - \tilde{\mathcal{A}}_{-E_{p}}\right) E_{p} \right] \biggr\}&=0\,,k=0,1,2,\ldots,n\,.\label{SubLeading}
\end{align}
\end{keyeqn}
Here we explicitly include the symmetry parameters since for $n \geq 2$ they can be traceless and therefore play an important role. Note that the $k=0$ and $k=1$ conditions of  \eqref{SubLeading} are trivial. We see that these conditions form a very neat tower structure, as desired. After discussing symmetries that depend on time, we will discuss the general solutions to these soft theorems, and present some examples.  

\paragraph{Linear-in-time symmetry ($\delta \pi = c t$)} We now turn our attention to symmetries that depend on time and start with the simplest time-dependent symmetry where there is no dependence on the spatial coordinates. We will then include additional powers of $x$. Note that at this stage we don't consider symmetries of this type that are non-linear in $t$ since these can never be a symmetry of the free theory (unless we are in the Lorentz-invariant limit which we will discuss later), as we discussed in the introduction. The analog of \eqref{WTIntermediate} is now  
\begin{align}
   \lim_{q\rightarrow 0} \biggl\{\partial_{q_0}\left[\frac{q^0+E_{q}}{2E_{q}}e^{i(q^0-E_{q})T}\mathcal{A}_{E_{q}}\delta(E_{q}+p^0) + \frac{-q^0+E_{q}}{2E_{q}}e^{i(q^0+E_{q})T}\mathcal{A}_{-E_{q}}\delta(-E_{q}+p^0) \right] \delta^3(\vec{q}+\vec{p}) \biggr\} = 0\,,
\end{align}
and by using the delta functions and taking the derivative we find
\begin{align}
\lim_{\vec{p}\rightarrow 0} \biggl\{ \frac{\tilde{\mathcal{A}}_{E_{p}}-\tilde{\mathcal{A}}_{-E_{p}}}{E_{p}}+iT(\tilde{\mathcal{A}}_{E_{p}}+\tilde{\mathcal{A}}_{-E_{p}}) \biggr\}=0\,.
\end{align}
We want this condition to hold for all $T$ so we actually have two conditions with the first being the Adler zero and the second a new condition:
\begin{keyeqn}
\begin{align}
\lim_{\vec{p}\rightarrow 0} \biggl\{\tilde{\mathcal{A}}_{E_{p}} + \tilde{\mathcal{A}}_{-E_{p}} \biggr\} & =0\,, \\
\lim_{\vec{p}\rightarrow 0} \biggl\{ \frac{\tilde{\mathcal{A}}_{E_{p}}-\tilde{\mathcal{A}}_{-E_{p}}}{E_{p}} \biggr\} & =0\,.
\end{align}
\end{keyeqn}
Again, it is comforting that we recover the Adler zero condition since it is a necessary condition for the closure of the algebra with the relevant commutator being
\begin{align}
[P_{0},C]  =A\,,
\end{align}
where $C$ is the generator of the $\delta \pi = c  t$ symmetry and $A$ is, as before, the generator of the $\delta \pi = a$ symmetry. Here the tower structure arises from derivatives with respect to $q^0$ acting on the exponential factor. 

\paragraph{General linear-in-time symmetry ($\delta \pi  = c_{i_{1} \ldots i_{n}} t x^{i_{1}} \ldots x^{i_{n}}$)} We now allow for dependence on the spatial coordinate too. The derivation here is very similar to what we have encountered above, but now with derivatives with respect to the spatial momentum acting on the momentum-conserving delta function. By taking care with derivatives and delta functions we find 
\begin{keyeqn}
\begin{align}
c_{i_{1} \ldots i_{n}} \lim_{\vec{p}\rightarrow 0} \biggl\{\partial_{p_{i_1}}\dots\partial_{p_{i_k}}\left[ \tilde{\mathcal{A}}_{E_{p}}+ \tilde{\mathcal{A}}_{-E_{p}} \right] \biggr\} & = 0 \,,\;\;\;k=0,1,2,\ldots,n \,,  \label{Leadingt1New} \\
c_{i_{1} \ldots i_{n}} \lim_{\vec{p}\rightarrow 0} \biggl\{\partial_{p_{i_1}}\dots\partial_{p_{i_k}}\left[\frac{\tilde{\mathcal{A}}_{E_{p}}-\tilde{\mathcal{A}}_{-E_{p}}}{E_{p}} \right] \biggr\} & =0 \,,\;\;\;k=0,1,2,\ldots,n \,, \label{SubLeadingtNew}
\end{align}
\end{keyeqn}
as the final form of the linear-in-time soft theorems. We note that any solutions to this system also solve the general spatial soft theorems for the same $n$. Indeed, \eqref{SubLeadingtNew} are stronger constraints than \eqref{SubLeading}. This is precisely our desired tower structure given that the corresponding symmetries are related by a time translation. We also note that structure of the amplitudes in \eqref{SubLeadingtNew}, namely $(\tilde{\mathcal{A}}_{E_{p}} - \tilde{\mathcal{A}}_{-E_{p}})/E_{p}$, is reminiscent of the shifted wavefunction coefficient used to derive soft theorems for flat-space wavefunction coefficients in \cite{Bittermann:2022nfh}. 

\subsection{Solutions to purely-spatial soft theorems  (\texorpdfstring{$\delta \pi = b_{i_{1} \ldots i_{n}}x^{i_{1}} \ldots x^{i_{n}}$)}{Lg} }

Let's now consider solutions to these conditions, initially concentrating on the time-independent symmetries with conditions \eqref{Leading} and \eqref{SubLeading}. We discuss solutions in terms of the soft amplitudes and note that finding hard on-shell amplitudes that are Bose symmetric and reduce to these soft amplitude solutions is still a non-trivial task which we tackle in Section \ref{sec:bootstrap}. The solutions split into two sets depending on whether the symmetry parameter is traceless or not. If it is traceful, then the \eqref{SubLeading} conditions require the amplitudes to be even in the soft energy, or of the form $\tilde{\mathcal{A}} = \mathcal{O}(p_1^{n})$ if they are odd in the soft energy. The conditions \eqref{Leading} then permit the terms that are odd in the soft energy, but those even in the soft energy must be of the form $\tilde{\mathcal{A}} = \mathcal{O}(p_1^{n+1})$. These solutions are shown in Figure \ref{fig:PurelySpatialTraceful}, where the solutions are expressed in terms of the minimal basis such that the dependence on the soft four-momentum only enters via its energy $E_{1}$ and contractions with other hard momenta $\vec{p}_{1} \cdot \vec{p}_{c}$ where here $a = 2 , \ldots n-1$. As we discussed in the introduction, invariance of the free theory requires this symmetry parameter to be traceless, unless we are interested in the Lorentz-invariant case where this parameter needs to contain a trace which then combines with symmetries that are non-linear in time to realise invariance of the free theory. We consider these solutions such that we can discuss the Lorentz-invariant limit below.
\begin{figure}[ht!]
    \centering    
\begin{minipage}{.45\textwidth}
\begin{tikzpicture} [scale=.65, every node/.style={scale=0.65}]
 \node  {$\Ccancel{1}$}[sibling distance = 1.9cm]
    child {node (10){$\Ccancel{E_{1}}$} 
    child {node (20){$\Ccancel{E_{1}^2}$}
    child {node (30){$\Ccancel{E_{1}^3}$}
    child {node (n0){$\Ccancel{E_{1}^n}$}
    child {node (n+10){$E_{1}^{n+1}$} edge from parent[draw=none]}
    child {node (n1){$\dots$} edge from parent[draw=none]}
    edge from parent[draw=none]}
    child {node (n-11){$E_{1}^{n-1}(\vec{p}_{1} \cdot \vec{p}_{c})$} 
    child {node { } edge from parent[draw=none]}
    child {node {$\ldots$} edge from parent[draw=none]} edge from parent[draw=none]} edge from parent[draw=none]}
    child {node (21){$\Ccancel{E_{1}^2(\vec{p}_{1} \cdot \vec{p}_{c})}$} 
     edge from parent[draw=none]} edge from parent[draw=none]}
    child {node { }
    child {node { } edge from parent[draw=none]}
    child {node { } edge from parent[draw=none]} edge from parent[draw=none]} edge from parent[draw=none]}
    child {node (01){$\Ccancel{\vec{p}_{1} \cdot \vec{p}_{c}}$} 
    child {node (11){$\Ccancel{E_{1} \vec{p}_{1} \cdot \vec{p}_{c}}$} edge from parent[draw=none]}
    child {node (02){$\Ccancel{(\vec{p}_{1} \cdot \vec{p}_{c})^2}$}
    child {node (12){$\Ccancel{E_{1}(\vec{p}_{1} \cdot \vec{p}_{c})^2}$}
    child {node (dots){$\cdots$} edge from parent[draw=none]}
    child {node (1n-1){$E_{1}(\vec{p}_{1} \cdot \vec{p}_{c})^{n-1}$}
    child {node {$\ldots$} edge from parent[draw=none]}
    child {node { } edge from parent[draw=none]}
    edge from parent[draw=none]} edge from parent[draw=none]}
    child {node (03){$\Ccancel{(\vec{p}_{1} \cdot \vec{p}_{c})^3}$}
    child {node { } edge from parent[draw=none]}
    child {node (0n){$\Ccancel{(\vec{p}_{1} \cdot \vec{p}_{c})^{n}}$}
   child {node (1n){$\dots$} 
   edge from parent[draw=none]}
    child {node (0n+1){$(\vec{p}_{1} \cdot \vec{p}_{c})^{n+1}$} edge from parent[draw=none]} 
    edge from parent[draw=none]} 
 edge from parent[draw=none]} edge from parent[draw=none]} edge from parent[draw=none]};
    \path (30) -- (n0) node [midway] {$\vdots$};
    \path (21) -- (n-11) node [midway] {$\vdots$};
    \path (12) -- (1n-1) node [midway] {$\vdots$};
    \path (03) -- (0n) node [midway] {$\vdots$};
\end{tikzpicture}
\end{minipage}
\begin{minipage}{.45\textwidth}
\begin{tikzpicture} [scale=.65, every node/.style={scale=0.65}]
 \node  {$\Ccancel{1}$}[sibling distance = 1.9cm]
    child {node (10){$\Ccancel{E_{1}}$} 
    child {node (20){$\Ccancel{E_{1}^2}$}
    child {node (30){$\Ccancel{E_{1}^3}$}
    child {node (n0){$E_{1}^n$} 
    child {node (n+10){$E_{1}^{n+1}$} edge from parent[draw=none]}
    child {node (n1){$\dots$} edge from parent[draw=none]}
    edge from parent[draw=none]}
    child {node (n-11){$\Ccancel{E_{1}^{n-1}(\vec{p}_{1} \cdot \vec{p}_{c})}$} 
    child {node { } edge from parent[draw=none]}
    child {node {$\ldots$} 
    edge from parent[draw=none]}
    edge from parent[draw=none]} edge from parent[draw=none]}
    child {node (21){$\Ccancel{E_{1}^2(\vec{p}_{1} \cdot \vec{p}_{c})}$} 
     edge from parent[draw=none]} edge from parent[draw=none]}
    child {node { }
    child {node { } edge from parent[draw=none]}
    child {node { } edge from parent[draw=none]} edge from parent[draw=none]} edge from parent[draw=none]}
    child {node (01){$\Ccancel{\vec{p}_{1} \cdot \vec{p}_c}$} 
    child {node (11){$\Ccancel{E_{1} \vec{p}_{1} \cdot \vec{p}_c}$} edge from parent[draw=none]}
    child {node (02){$\Ccancel{(\vec{p}_{1} \cdot \vec{p}_{c})^2}$}
    child {node (12){$\Ccancel{E_{1}(\vec{p}_{1} \cdot \vec{p}_{c})^2}$}
    child {node (dots){$\cdots$} edge from parent[draw=none]}
    child {node (1n-1){$E_{1}(\vec{p}_{1} \cdot \vec{p}_a)^{n-1}$} 
    child {node {$\ldots$} edge from parent[draw=none]}
    child {node { } edge from parent[draw=none]}
    edge from parent[draw=none]} edge from parent[draw=none]}
    child {node (03){$\Ccancel{(\vec{p}_{1} \cdot \vec{p}_{c})^3}$}
    child {node { } edge from parent[draw=none]}
    child {node (0n){$\Ccancel{(\vec{p}_{1} \cdot \vec{p}_{c})^{n}}$} 
    child {node (1n){$\dots$} 
   edge from parent[draw=none]}
    child {node (0n+1){$(\vec{p}_{1} \cdot \vec{p}_{c})^{n+1}$} edge from parent[draw=none]} 
    edge from parent[draw=none]} 
 edge from parent[draw=none]} edge from parent[draw=none]} edge from parent[draw=none]};
    \path (30) -- (n0) node [midway] {$\vdots$};
    \path (21) -- (n-11) node [midway] {$\vdots$};
    \path (12) -- (1n-1) node [midway] {$\vdots$};
    \path (03) -- (0n) node [midway] {$\vdots$};
\end{tikzpicture}
\end{minipage}
\caption{These figures depict the solutions to the purely-spatial soft theorems, \eqref{Leading} and \eqref{SubLeading}, for which the symmetry parameter is traceful i.e. $\delta_{i_{1} i_{2}} b_{i_{1} i_{2} \ldots i_{n}} \neq 0$. The LHS solutions correspond to even $n$, while the RHS correspond to odd $n$.}
 \label{fig:PurelySpatialTraceful}
\end{figure}
If the symmetry parameter is traceless then additional solutions are allowed. The conditions \eqref{SubLeading} now admit all soft amplitudes since any amplitude that is even in the soft energy trivially cancels within the round brackets, while those odd in the soft energy survive in the round brackets but then become even in the soft energy when multiplied by $E_{p}$ and such terms are always allowed if the parameter is traceless since taking the derivatives and soft limit will always yield at least one copy of $\delta_{ij}$ in every non-trivial term. The conditions \eqref{Leading} then trivially allow all terms that are odd in the soft energy, while terms that are even in the soft energy are allowed thanks to the traceless condition as long as they are not zeroth order. Terms that are zeroth order must be of the form $\tilde{\mathcal{A}} = \mathcal{O}(p_1^{n+1})$. These solutions are shown in Figure \ref{fig:GenerousPurelySpatialTraceless}. We see how terms that are energy dependent pass the soft theorems with ease. This makes sense from a Lagrangian point of view since they arise from time derivatives which are trivially invariant under the purely-spatial symmetries. 
\begin{figure}[ht!]
    \centering    
\begin{tikzpicture} [scale=.7, every node/.style={scale=0.7}]
 \node  {$\Ccancel{1}$}[sibling distance = 2cm]
    child {node (10){$E_{1}$} 
    child {node (20){$E_{1}^2$}
    child {node (30){$E_{1}^3$}
    child {node (n0){$E_{1}^n$}
    child {node (n+10){$E_{1}^{n+1}$} edge from parent[draw=none]}
    child {node (n1){$\dots$} edge from parent[draw=none]}
    edge from parent[draw=none]}
    child {node (n-11){$E_{1}^{n-1}(\vec{p}_{1} \cdot \vec{p}_{c})$} 
    child {node { } edge from parent[draw=none]}
    child {node {$\ldots$} 
    edge from parent[draw=none]}
    edge from parent[draw=none]} edge from parent[draw=none]}
    child {node (21){$E_{1}^2(\vec{p}_{1} \cdot \vec{p}_{c})$} 
     edge from parent[draw=none]} edge from parent[draw=none]}
    child {node { }
    child {node { } edge from parent[draw=none]}
    child {node { } edge from parent[draw=none]} edge from parent[draw=none]} edge from parent[draw=none]}
    child {node (01){$\Ccancel{\vec{p}_{1} \cdot \vec{p}_{c}}$} 
    child {node (11){$E_{1} (\vec{p}_{1} \cdot \vec{p}_{c})$} edge from parent[draw=none]}
    child {node (02){$\Ccancel{(\vec{p}_{1} \cdot \vec{p}_{c})^2}$}
    child {node (12){$E_{1}(\vec{p}_{1} \cdot \vec{p}_{c})^2$}
    child {node (dots){$\cdots$} edge from parent[draw=none]}
    child {node (1n-1){$E_{1}(p_{1} \cdot \vec{p}_{c})^{n-1}$}
    child {node {$\ldots$} edge from parent[draw=none]}
    child {node { } edge from parent[draw=none]}
    edge from parent[draw=none]} edge from parent[draw=none]}
    child {node (03){$\Ccancel{(\vec{p}_{1} \cdot \vec{p}_{c})^3}$}
    child {node { } edge from parent[draw=none]}
    child {node (0n){$\Ccancel{(\vec{p}_{1} \cdot \vec{p}_{c})^{n}}$}
     child {node (1n){$\dots$} 
   edge from parent[draw=none]}
    child {node (0n+1){$(\vec{p}_{1} \cdot \vec{p}_{c})^{n+1}$} edge from parent[draw=none]} 
    edge from parent[draw=none]} 
 edge from parent[draw=none]} edge from parent[draw=none]} edge from parent[draw=none]};
    \path (30) -- (n0) node [midway] {$\vdots$};
    \path (21) -- (n-11) node [midway] {$\vdots$};
    \path (12) -- (1n-1) node [midway] {$\vdots$};
    \path (03) -- (0n) node [midway] {$\vdots$};
\end{tikzpicture}
\caption{This figure depicts the solutions to the purely-spatial soft theorems, \eqref{Leading} and \eqref{SubLeading}, for which the symmetry parameter is traceless i.e. $\delta_{i_{1} i_{2}} b_{i_{1} i_{2} \ldots i_{n}} = 0$. These solutions are valid for both even and odd $n$.}
\label{fig:GenerousPurelySpatialTraceless}
\end{figure}


\subsection{Solutions to linear-in-time soft theorems (\texorpdfstring{$\delta \pi = c_{i_{1} \ldots i_{n}} t x^{i_{1}} \ldots x^{i_{n}}$}{Lg})}
We now consider solutions to the soft theorems given in \eqref{Leadingt1New} and \eqref{SubLeadingtNew}. Again we consider the two cases where the symmetry parameter is traceful or traceless. If it is traceful, then all $\mathcal{O}(p^{n+2})$ solutions are allowed, as are $\mathcal{O}(p^{n+1})$ ones if they are even in the soft energy. These solutions are shown in Figure \ref{fig:LinearinTimeTraceful}.
\begin{figure}[ht!]
    \centering    
\begin{minipage}{.45\textwidth}
\begin{tikzpicture} [scale=.65, every node/.style={scale=0.65}]
 \node  {$\Ccancel{1}$}[sibling distance = 1.9cm]
    child {node (10){$\Ccancel{E_{1}}$} 
    child {node (20){$\Ccancel{E_{1}^2}$}
    child {node (30){$\Ccancel{E_{1}^3}$}
    child {node (n0){$\Ccancel{E_{1}^n}$}
    child {node (n+10){$\Ccancel{E_{1}^{n+1}}$} 
    edge from parent[draw=none]}
    child {node (n1){$E_{1}^{n}(\vec{p}_{1} \cdot \vec{p}_{c})$} edge from parent[draw=none]}
    edge from parent[draw=none]}
    child {node (n-11){$\Ccancel{E_{1}^{n-1}(\vec{p}_{1} \cdot \vec{p}_{c})}$} 
    child {node { } edge from parent[draw=none]}
    child {node {$\ldots$} edge from parent[draw=none]} edge from parent[draw=none]} edge from parent[draw=none]}
    child {node (21){$\Ccancel{E_{1}^2(\vec{p}_{1} \cdot \vec{p}_{c})}$} 
     edge from parent[draw=none]} edge from parent[draw=none]}
    child {node { }
    child {node { } edge from parent[draw=none]}
    child {node { } edge from parent[draw=none]} edge from parent[draw=none]} edge from parent[draw=none]}
    child {node (01){$\Ccancel{\vec{p}_{1} \cdot \vec{p}_{c}}$} 
    child {node (11){$\Ccancel{E_{1} (\vec{p}_{1} \cdot \vec{p}_{c})}$} edge from parent[draw=none]}
    child {node (02){$\Ccancel{(\vec{p}_{1} \cdot \vec{p}_{c})^2}$}
    child {node (12){$\Ccancel{E_{1}(\vec{p}_{1} \cdot \vec{p}_{c})^2}$}
    child {node (dots){$\cdots$} edge from parent[draw=none]}
    child {node (1n-1){$\Ccancel{E_{1}(\vec{p}_{1} \cdot \vec{p}_{c})^{n-1}}$}
    child {node {$\ldots$} edge from parent[draw=none]}
    child {node { } edge from parent[draw=none]}
    edge from parent[draw=none]} edge from parent[draw=none]}
    child {node (03){$\Ccancel{(\vec{p}_{1} \cdot \vec{p}_{c})^3}$}
    child {node { } edge from parent[draw=none]}
    child {node (0n){$\Ccancel{(\vec{p}_{1} \cdot \vec{p}_{c})^{n}}$}
    child {node (1n){$\Ccancel{E_{1}(\vec{p}_{1} \cdot \vec{p}_{c})^n}$} 
    edge from parent[draw=none]}
    child {node (0n+1){$(\vec{p}_{1} \cdot \vec{p}_{c})^{n+1}$} 
    edge from parent[draw=none]} 
 edge from parent[draw=none]} 
 edge from parent[draw=none]} edge from parent[draw=none]} edge from parent[draw=none]};
    \path (30) -- (n0) node [midway] {$\vdots$};
    \path (21) -- (n-11) node [midway] {$\vdots$};
    \path (12) -- (1n-1) node [midway] {$\vdots$};
    \path (03) -- (0n) node [midway] {$\vdots$};
\end{tikzpicture}
\end{minipage}
\begin{minipage}{.45\textwidth}
\begin{tikzpicture} [scale=.65, every node/.style={scale=0.65}]
 \node  {$\Ccancel{1}$}[sibling distance = 1.9cm]
    child {node (10){$\Ccancel{E_{1}}$} 
    child {node (20){$\Ccancel{E_{1}^2}$}
    child {node (30){$\Ccancel{E_{1}^3}$}
    child {node (n0){$\Ccancel{E_{1}^n}$}
    child {node (n+10){$E_{1}^{n+1}$}
    edge from parent[draw=none]}
    child {node (n1){$\Ccancel{E_{1}^{n}(\vec{p}_{1} \cdot \vec{p}_{c})}$} edge from parent[draw=none]}
    edge from parent[draw=none]}
    child {node (n-11){$\Ccancel{E_{1}^{n-1}(\vec{p}_{1} \cdot \vec{p}_{c})}$} 
    child {node { } edge from parent[draw=none]}
    child {node {$\ldots$} edge from parent[draw=none]} edge from parent[draw=none]} edge from parent[draw=none]}
    child {node (21){$\Ccancel{E_{1}^2(\vec{p}_{1} \cdot \vec{p}_{c})}$} 
     edge from parent[draw=none]} edge from parent[draw=none]}
    child {node { }
    child {node { } edge from parent[draw=none]}
    child {node { } edge from parent[draw=none]} edge from parent[draw=none]} edge from parent[draw=none]}
    child {node (01){$\Ccancel{\vec{p}_{1} \cdot \vec{p}_{c}}$} 
    child {node (11){$\Ccancel{E_{1} (\vec{p}_{1} \cdot \vec{p}_{c})}$} edge from parent[draw=none]}
    child {node (02){$\Ccancel{(\vec{p}_{1} \cdot \vec{p}_{c})^2}$}
    child {node (12){$\Ccancel{E_{1}(\vec{p}_{1} \cdot \vec{p}_{c})^2}$}
    child {node (dots){$\cdots$} edge from parent[draw=none]}
    child {node (1n-1){$\Ccancel{E_{1}(\vec{p}_{1} \cdot \vec{p}_{c})^{n-1}}$}
    child {node {$\ldots$} edge from parent[draw=none]}
    child {node { } edge from parent[draw=none]}
    edge from parent[draw=none]} edge from parent[draw=none]}
    child {node (03){$\Ccancel{(\vec{p}_{1} \cdot \vec{p}_{c})^3}$}
    child {node { } edge from parent[draw=none]}
    child {node (0n){$\Ccancel{(\vec{p}_{1} \cdot \vec{p}_{c})^{n}}$}
   child {node (1n){$\Ccancel{E_{1}(\vec{p}_{1} \cdot \vec{p}_{c})^n}$} 
   edge from parent[draw=none]}
    child {node (0n+1){$(\vec{p}_{1} \cdot \vec{p}_{c})^{n+1}$}
    edge from parent[draw=none]} 
    edge from parent[draw=none]} 
 edge from parent[draw=none]} edge from parent[draw=none]} edge from parent[draw=none]};
    \path (30) -- (n0) node [midway] {$\vdots$};
    \path (21) -- (n-11) node [midway] {$\vdots$};
    \path (12) -- (1n-1) node [midway] {$\vdots$};
    \path (03) -- (0n) node [midway] {$\vdots$};
\end{tikzpicture}
\end{minipage}
\caption{These figures depict the solutions to the linear-in-time soft theorems, \eqref{Leadingt1New} and \eqref{SubLeadingtNew}, for which the symmetry parameter is traceful i.e. $\delta_{i_{1} i_{2}} c_{i_{1} i_{2} \ldots i_{n}} \neq 0$. The LHS solutions correspond to even $n$, while the RHS ones correspond to odd $n$.}
 \label{fig:LinearinTimeTraceful}
\end{figure}
If the symmetry parameter is traceless then the solution set is larger and is shown in Figure \ref{fig:NewLinearInTimeTraceless}.
\begin{figure}[ht!]
    \centering    
\begin{tikzpicture} [scale=.65, every node/.style={scale=0.65}]
 \node  {$\Ccancel{1}$}[sibling distance = 1.9cm]
    child {node (10){$\Ccancel{E_{1}}$} 
    child {node (20){$E_{1}^2$}
    child {node (30){$E_{1}^3$}
    child {node (n0){$E_{1}^n$}
    child {node (n+10){$E_{1}^{n+1}$}
    child {node (n+20){$E_{1}^{n+2}$} 
    edge from parent[draw=none]}
    child {node {$\dots$} 
    edge from parent[draw=none]}
    edge from parent[draw=none]}
    child {node (n1){$\dots$} edge from parent[draw=none]}
    edge from parent[draw=none]}
    child {node (n-11){$E_{1}^{n-1}(\vec{p}_{1} \cdot \vec{p}_{c})$} 
    child {node { } edge from parent[draw=none]}
    child {node {$\ldots$} edge from parent[draw=none]} edge from parent[draw=none]} edge from parent[draw=none]}
    child {node (21){$E_{1}^2(\vec{p}_{1} \cdot \vec{p}_{c})$} 
     edge from parent[draw=none]} edge from parent[draw=none]}
    child {node { }
    child {node { } edge from parent[draw=none]}
    child {node { } edge from parent[draw=none]} edge from parent[draw=none]} edge from parent[draw=none]}
    child {node (01){$\Ccancel{\vec{p}_{1} \cdot \vec{p}_{c}}$} 
    child {node (11){$\Ccancel{E_{1} (\vec{p}_{1} \cdot \vec{p}_{c}})$} edge from parent[draw=none]}
    child {node (02){$\Ccancel{(\vec{p}_{1} \cdot \vec{p}_{c})^2}$}
    child {node (12){$\Ccancel{E_{1}(\vec{p}_{1} \cdot \vec{p}_{c})^2}$}
    child {node (dots){$\cdots$} edge from parent[draw=none]}
    child {node (1n-1){$\Ccancel{E_{1}(\vec{p}_{1} \cdot \vec{p}_{c})^{n-1}}$}
    child {node {$\ldots$} edge from parent[draw=none]}
    child {node { } edge from parent[draw=none]}
    edge from parent[draw=none]} edge from parent[draw=none]}
    child {node (03){$\Ccancel{(\vec{p}_{1} \cdot \vec{p}_{c})^3}$}
    child {node { } edge from parent[draw=none]}
    child {node (0n){$\Ccancel{(\vec{p}_{1} \cdot \vec{p}_{c})^{n}}$}
   child {node (1n){$\Ccancel{E_{1}(\vec{p}_{1} \cdot \vec{p}_{c})^n}$} 
   edge from parent[draw=none]}
   child {node (0n+1){$(\vec{p}_{1} \cdot \vec{p}_{c})^{n+1}$}
   child {node {$\dots$} 
   edge from parent[draw=none]} 
   child {node (0n+2){$(\vec{p}_{1} \cdot \vec{p}_{c})^{n+2}$} edge from parent[draw=none]}
   edge from parent[draw=none]} 
   edge from parent[draw=none]} 
 edge from parent[draw=none]} edge from parent[draw=none]} edge from parent[draw=none]};
    \path (30) -- (n0) node [midway] {$\vdots$};
    \path (21) -- (n-11) node [midway] {$\vdots$};
    \path (12) -- (1n-1) node [midway] {$\vdots$};
    \path (03) -- (0n) node [midway] {$\vdots$};
\end{tikzpicture}
\caption{This figure depicts solutions to the linear-in-time soft theorems, \eqref{Leadingt1New} and \eqref{SubLeadingtNew}, for which the symmetry parameter is traceless i.e. $\delta_{i_{1} i_{1}} c_{i_{1} i_{2} \ldots i_{n}} = 0$. These solutions are valid for both even and odd $n$.}
 \label{fig:NewLinearInTimeTraceless}
\end{figure}

\subsection{Lorentz-invariant limit (\texorpdfstring{$\delta \pi = d_{\mu_{1} \ldots \mu_{n}} x^{\mu_{1}} \ldots x^{\mu_{n}}$}{Lg})}

Before discussing these solutions in more detail and providing some examples, let's see how we can recover the known Lorentz-invariant results derived in \cite{Cheung:2014dqa,Cheung:2016drk} where the symmetry $\delta \pi = d_{\mu_{1} \ldots \mu_{n}} x^{\mu_{1}} \ldots x^{\mu_{n}}$ requires amplitudes to scale as $\tilde{\mathcal{A}} = \mathcal{O}(p^{n+1})$. Here we are assuming that the amplitude is Lorentz invariant and so is a function of Lorentz-invariant contractions ($p_{a} \cdot p_{b} = \eta_{\mu\nu} p_{a}^{\mu}p_{b}^{\nu}$). To recover this result we need to combine the solutions for the purely-spatial symmetries, the solutions for the linear-in-time symmetries, but also include a new set of solutions corresponding to symmetries that are non-linear in time. In each case we need the symmetry parameters to contain the trace since invariance of the free theory now dictates $\eta^{\mu_{1}\mu_{2}} d_{\mu_{1} \mu_{2} \mu_{3} \ldots \mu_{n}} = 0$. Our first task is therefore to derive the soft theorems corresponding to symmetries that are non-linear in time. First consider the $\delta \pi = e_{i} t^2 x^{i}$ symmetry where the analog of \eqref{WTIntermediate} is
\begin{align}
   \lim_{q\rightarrow 0} \biggl\{\partial^2_{q_0}\left[\frac{q^0+E_{p}}{2E_{p}}e^{i(q^0-E_{p})T}\mathcal{A}_{E_{p}}\delta(E_{p}+p^0) + \frac{-q^0+E_{p}}{2E_{p}}e^{i(q^0+E_{p})T}\mathcal{A}_{-E_{p}}\delta(-E_{p}+p^0) \right] \partial_{p_{i}}\delta^3(\vec{q}+\vec{p}) \biggr\} = 0\,.
\end{align}
By computing the derivatives and taking the soft limit, we find that the necessary conditions are the same as \eqref{Leadingt1New} and \eqref{SubLeadingtNew} with $n=1$. This is also true for the more general non-linear-in-time symmetries: the soft theorem constraints are always given by \eqref{Leadingt1New} and \eqref{SubLeadingtNew} with the powers of the spatial coordinates matched in the two cases. To extract the Lorentz-invariant solutions we therefore need to combine the solutions we have already discussed above. We first note the solutions for the purely-spatial symmetries rule out all $(\vec{p}_{1} \cdot \vec{p}_{c})^{m}$ terms with $m \leq n$, as shown in Figure \ref{fig:PurelySpatialTraceful}. When the amplitudes are Lorentz invariant, this would also rule out any energy-dependent terms for which the total number of energy and momentum factors is also given by $m$. The relevant solutions are then those that scale as $ \tilde{\mathcal{A}} = \mathcal{O}(p^{n+1})$. These solutions then trivially solve the linear-in-time soft theorems (noting that the relevant symmetry is $\delta \pi = c_{i_{1} \ldots i_{n-1}} t x^{i_{1}} \ldots x^{i_{n-1}}$). We therefore see that our soft theorems recover the known Lorentz-invariant results of \cite{Cheung:2014dqa,Cheung:2016drk}.

\subsection{Examples}

Here we consider some examples of interactions that are invariant under some of the symmetries we have considered and check that the corresponding amplitudes satisfy our soft theorems. 

\paragraph{Example 1: $\dot{\pi}^4$} This interaction is clearly invariant under all of the purely-spatial symmetries and therefore the corresponding amplitudes should satisfy all the soft theorems in \eqref{Leading} and \eqref{SubLeading}. This interaction is not invariant under any of the linear-in-time symmetries and therefore we would expect the amplitudes to not satisfy any of the soft theorems in \eqref{Leadingt1New} and \eqref{SubLeadingtNew}. The contact four-point scattering amplitude written in terms of the minimal basis is $\tilde{\mathcal{A}}_{E_{1}} =- E_{1} E_{2} E_{3} (E_{1} + E_{2}  + E_{3})$, so we have 
\begin{align}
\tilde{\mathcal{A}}_{E_{1}} + \tilde{\mathcal{A}}_{-E_{1}} &= - 2 E_{1}^2 E_{2}E_{3} \,, \\
(\tilde{\mathcal{A}}_{E_{1}} - \tilde{\mathcal{A}}_{-E_{1}})E_{1} &= - 2 E_{1}^2 E_{2}E_{3} (E_{2} + E_{3}) \,.
\end{align}
First consider the soft theorems in \eqref{Leading}. The conditions with $k \neq 2$ are trivially satisfied, while for $k=2$ this condition is satisfied if $\delta_{i_{1} i_{2}} b_{i_{1}i_{2}i_{3} \ldots i_{n}} = 0$ i.e. if the symmetry parameters are traceless. Similarly, the only non-trivial condition out of those in \eqref{SubLeading} is $k=2$ and again this soft theorem requires the symmetry parameters to be traceless. As we have discussed a number of times, the free theory is only invariant under these symmetries if the parameters are traceless so it is comforting that we find such traceless conditions as being necessary at the level of the soft theorem. Here we see that the tower structure is crucial since for $n > 2$ it is always the $k=2$ condition that is important. For the linear-in-time soft theorems, the conditions \eqref{Leadingt1New} are satisfied, however those in \eqref{SubLeadingtNew} are not. Indeed, we have 
\begin{align}
\frac{\tilde{\mathcal{A}}_{E_{1}} - \tilde{\mathcal{A}}_{-E_{1}}}{E_{1}} &= - 2 E_{2}E_{3}(E_{2}+E_{3}) \,,
\end{align}
which fails the $k=0$ condition of \eqref{SubLeadingtNew}. This condition is there for all $n$ by the tower structure so the amplitude does not satisfy any of the linear-in-time soft theorems, as expected. 

\paragraph{Example 2: $\ddot{\pi}^4$} This interaction is also invariant under all of the purely-spatial symmetries, but in contrast to the previous example, it is also invariant under all linear-in-time symmetries. The corresponding amplitudes should therefore satisfy all of our soft theorems. The four-point amplitude in the minimal basis is $\tilde{\mathcal{A}}_{E_{1}} = E_{1}^2 E_{2}^2 E_{3}^2 (E_{1}+E_{2}+E_{3})^2$, and therefore 
\begin{align}
\tilde{\mathcal{A}}_{E_{1}} + \tilde{\mathcal{A}}_{-E_{1}} &= 2 E_{1}^2 E_{2}^2 E_{3}^2 [ E_{1}^2 + (E_{2}+E_{3})^2]  \,, \\
(\tilde{\mathcal{A}}_{E_{1}} - \tilde{\mathcal{A}}_{-E_{1}})E_{1} &= 4 E_{1}^4 E_{2}^2 E_{3}^2 (E_{2}+E_{3})  \,.
\end{align}
We therefore again see that all of the conditions in \eqref{Leading} and \eqref{SubLeading} are satisfied as long as the symmetry parameters are traceless. We then also have 
\begin{align}
\frac{\tilde{\mathcal{A}}_{E_{1}} - \tilde{\mathcal{A}}_{-E_{1}}}{E_1} &= 4 E_{1}^2 E_{2}^2 E_{3}^2 (E_{2}+E_{3})  \,,
\end{align}
which satisfies all the conditions in \eqref{SubLeadingtNew} as long as the symmetry parameters are traceless. Note that the overall factor of $E_{1}^2$ is enough to guarantee that all soft theorems are satisfied. 

\paragraph{Example 3: $\dot{\pi}^2 (\partial_{i}\partial_{j} \pi)^2$} This vertex is invariant under the $n=1$ purely-spatial symmetry, but is not invariant under the $n \geq 2$ ones. It is also not invariant under any of the linear-in-time soft theorems. The four-point amplitude without energy and momentum conservation imposed is 
\begin{align}
\mathcal{A} = [E_{1}E_{2} (\vec{p}_{3} \cdot \vec{p}_{4})^2 +E_{1}E_{3} (\vec{p}_{2} \cdot \vec{p}_{4})^2 +E_{1}E_{4} (\vec{p}_{1} \cdot \vec{p}_{3})^2 + E_{2}E_{3} (\vec{p}_{1} \cdot \vec{p}_{4})^2 \nonumber \\ + E_{2}E_{4} (\vec{p}_{1} \cdot \vec{p}_{3})^2 + E_{3}E_{4} (\vec{p}_{1} \cdot \vec{p}_{2})^2] \delta(E_{1}+E_{2}+E_{3}+E_{4}) \delta^{(3)}(\vec{p}_{1}+\vec{p}_{2}+\vec{p}_{3}+\vec{p}_{4}).
\end{align}
The expression for the minimal basis amplitude is somewhat messy but the combinations we care about take a simplified form. For example, we have 
\begin{align}
\tilde{\mathcal{A}}_{E_{1}} +  \tilde{\mathcal{A}}_{-E_{1}} = & -8[(\vec{p}_{1} \cdot \vec{p}_{2})^2 (E_{1}^2 + E_{3}^2) - 2 E_{1}^2  (\vec{p}_{1} \cdot \vec{p}_{2}) E_{3}( 3 E_{2} + E_{3} ) \nonumber \\ &  + (\vec{p}_{1} \cdot \vec{p}_{3})^2 (E_{1}^2 + E_{2}^2) - 2 E_{1}^2  (\vec{p}_{1} \cdot \vec{p}_{3}) E_{2}( 3 E_{3} + E_{2} ) \nonumber \\ &  + E_{1}^4 (E_{2}^2 + E_{3}^2 + E_{2} E_{3}) + 2 (\vec{p}_{1} \cdot \vec{p}_{2}) (\vec{p}_{1} \cdot \vec{p}_{3}) (E_{1}^2 - E_{2} E_{3})] \,,
\end{align}
and given that this combination is at least quadratic in the soft momentum it is simple to see that the $n=1$ purely-spatial soft theorem is satisfied. It is also easy to see that for $n \geq 2$ the soft theorems are not satisfied since
\begin{align} 
   b_{ij} \lim_{\vec{p}\rightarrow 0}\partial_{p_{i}}\partial_{p_{j}}(\tilde{\mathcal{A}}_{E_{1}}+\tilde{\mathcal{A}}_{-E_{1}}) \supset -16  b_{ij} p^{2}_{i}p^{2}_{j} E_{3}^2 \neq 0\,.
\end{align}
This reproduces the correct results for the purely-spatial soft theorems since for $n=1$ the conditions in \eqref{SubLeading} are trivial. For the linear-in-time soft theorems we note that  
\begin{align}
\tilde{\mathcal{A}}_{E_{1}} \supset E_{1} E_{2} E_{3} (E_{2} + E_{3}) (E_{2}^2 + E_{2} E_{3} + E_{3}^2)\,,
\end{align}
and therefore 
\begin{align}
\frac{\tilde{\mathcal{A}}_{E_{1}} - \tilde{\mathcal{A}}_{-E_{1}}}{E_{1}}   \supset 2 E_{2} E_{3} (E_{2} + E_{3}) (E_{2}^2 + E_{2} E_{3} + E_{3}^2)\,.
\end{align}
It follows that the $k=0$ condition of \eqref{SubLeadingtNew} is not satisfied. Our soft theorems therefore reproduce exactly what we would expect. 

\paragraph{Example 4: $(\partial_i \pi)^2 [g_{1}(\partial_i^2 \pi)^2  + g_{2}(\partial_i \partial_j \pi)^2 ]$} In this case the four-point amplitude is 
\begin{align}
\mathcal{A} =  \biggl\{ \vec{p}_{1} \cdot \vec{p}_{2} \left[g_{1} E_{3}^2E_{4}^2 + g_{2} (\vec{p}_{3} \cdot \vec{p}_{4})^2 \right] + \text{perms} \biggr\} \delta(E_{1}+E_{2}+E_{3}+E_{4}) \delta^{(3)}(\vec{p}_{1}+\vec{p}_{2}+\vec{p}_{3}+\vec{p}_{4})\,.
\end{align}
The combination $\tilde{\mathcal{A}}_{E_{1}} - \tilde{\mathcal{A}}_{-E_{1}}$ vanishes since the minimal basis amplitude is even in the soft energy $E_{1}$ for any $g_{1}$ and $g_{2}$. It follows that the $n=0$ linear-in-time soft theorems are satisfied which makes sense since these vertices enjoy the corresponding symmetry for any $g_{1}$ and $g_{2}$. For the spatial soft theorems, the  $k=1$ condition of \eqref{Leading} yields
\begin{align}
    \lim_{\vec{p}\rightarrow 0}\partial_{p_{i}}(\tilde{\mathcal{A}}_{E_{1}} +  \tilde{\mathcal{A}}_{-E_{1}}) = 8 (g_{1}+g_{2})p^{2}_{i} E_{3}^2(2E_{2}+E_{3}) +8 (g_{1}+g_{2})p^{3}_{i} E_{2}^2(2E_{3}+E_{2})\,, 
\end{align}
and so this expression only vanishes for $g_{1} = -g_{2}$ and indeed the interaction is only invariant under the $\delta \pi = b_{i}x^{i}$ symmetry if we have this tuning of the couplings. Then according to the tower structure, the amplitude would also satisfy the $n=1$ linear-in-time soft theorem and indeed for $g_{1} =  -g_{2}$ the vertex enjoys the corresponding symmetry. The $k=2$ condition for the purely-spatial soft theorem reads
\begin{align} 
   b_{ij} \lim_{\vec{p}\rightarrow 0}\partial_{p_{i}}\partial_{p_{j}}(\tilde{\mathcal{A}}_{E_{1}}+\tilde{\mathcal{A}}_{-E_{1}}) \supset -24 g_{1} b_{ij} p^{2}_{i}p^{2}_{j} E_{3}^2 \neq 0\,.
\end{align}
Therefore the higher-order soft theorems are not satisfied. As we will show in Appendix \ref{app:cosets}, this vertex is indeed a Wess-Zumino term for the $\delta\pi = c_i tx^i$ and $\delta \pi = b_i x^i$ symmetries when picking out $g_1 = - g_2$ so our soft theorem analysis agrees with an off-shell analysis.

\paragraph{Example 5: $(\partial_i \dot{\pi})^2 [g_{3}(\partial_i^2 \dot{\pi})^2  + g_{4}(\partial_i \partial_j \dot{\pi})^2 ]$} In this case the four-point amplitude is 
\begin{align}
\mathcal{A} = E_{1}E_{2} E_{3}E_{4} \biggl\{ \vec{p}_{1} \cdot \vec{p}_{2} \left[g_{3} E_{3}^2E_{4}^2 + g_{4} (\vec{p}_{3} \cdot \vec{p}_{4})^2 \right] + \text{perms} \biggr\} \delta(E_{1}+E_{2}+E_{3}+E_{4}) \delta^{(3)}(\vec{p}_{1}+\vec{p}_{2}+\vec{p}_{3}+\vec{p}_{4})\,.
\end{align}
The expressions for $\tilde{\mathcal{A}}_{E_{1}} +  \tilde{\mathcal{A}}_{-E_{1}}$ and $\tilde{\mathcal{A}}_{E_{1}} -  \tilde{\mathcal{A}}_{-E_{1}}$ are again quite messy, however thanks to the overall factor of $E_{1}E_{2} E_{3}E_{4}$ we know that $\tilde{\mathcal{A}}_{E_{1}} +  \tilde{\mathcal{A}}_{-E_{1}}$ is at least quadratic in the soft energy and so all conditions in \eqref{Leading} are satisfied if the symmetry parameters are traceless. The overall factor of $E_{1}$ also ensures that all \eqref{SubLeading} conditions are satisfied. This makes sense from the symmetry point of view since for any $g_{3}$ and $g_{4}$ we clearly have invariance under all purely-spatial symmetries. Now consider the linear-in-time symmetries. The combination $\tilde{\mathcal{A}}_{E_{1}} -  \tilde{\mathcal{A}}_{-E_{1}}$ contains a term linear in $E_{1}$ with a coefficient that is linear in the soft momentum. It follows that the $k=0$ condition of \eqref{SubLeadingtNew} is satisfied. We also have 
\begin{align}
\lim_{\vec{p}\rightarrow 0} \biggl\{ \partial_{p_{i}}\left[\frac{\tilde{\mathcal{A}}_{E_{1}}-\tilde{\mathcal{A}}_{-E_{1}}}{E_{1}} \right] \biggr\} = 8 (g_{3}+g_{4})p^{2}_{i} E_{3}^2(2E_{2}+E_{3}) +8 (g_{3}+g_{4})p^{3}_{i} E_{2}^2(2E_{3}+E_{2})\,,
\end{align}
and so the $k=1$ condition in \eqref{SubLeadingtNew} is satisfied if $g_{3} = - g_{4}$. We have checked that with this tuning between the couplings the vertices do indeed enjoy the $n=1$ linear-in-time symmetry (since $(\partial_i^2 \dot{\pi})^2  - (\partial_i \partial_j \dot{\pi})^2$ is a total derivative) and in Appendix \ref{app:cosets} we show that this is actually a Wess-Zumino term for the corresponding symmetry. These vertices do not enjoy the higher-order linear-in-time symmetries and indeed the corresponding soft theorems are not satisfied.  

\paragraph{Example 6: $\dot{\pi}^4$ at six-points} So far we have concentrated on four-point amplitudes so let us end with an example at higher-points. We have already checked that the four-point amplitude for this vertex satisfies all the purely-spatial soft theorems. The relevant amplitude at six-points is a sum of $10$ channels but first consider the one that is singular when $(p^{\mu}_{1}+p^{\mu}_{2}+p^{\mu}_{3})^2 \rightarrow 0$. This propagator is $\mathcal{O}(1)$ in the soft limit so let's consider the other factors. The amplitude is at least linear in $E_{1}$ and therefore $\tilde{\mathcal{A}}_{E_{1}}+\tilde{\mathcal{A}}_{-E_{1}}$ is at least quadratic in $E_{1}$ and therefore all \eqref{Leading} conditions are satisfied if the symmetry parameters are traceless. The overall factor of $E_{1}$ also ensures that the \eqref{SubLeading} conditions are satisfied. For the same reason, the $k=0$ condition in \eqref{SubLeadingtNew} is not satisfied. The other channels work out in the same way. The soft theorem analysis is therefore consistent with an off-shell symmetry analysis.   

We have therefore checked that our soft theorems yield the expected results for a number of examples. In the following section we will further check the validity of these soft theorems by bootstrapping amplitudes that solve the conditions we have derived in this section, followed by comparing them with what we get from the coset construction.

\section{Soft bootstrap} \label{sec:bootstrap}

Having derived soft theorems and checked their validity against a few examples in the previous section, in this section we perform a more thorough analysis by using our soft theorems to derive the leading (in a momentum expansion) invariant quartic (and sextic) vertices for a number of different symmetries. We will then compare our results to what we would expect from a Lagrangian analysis by making use of the coset construction. We might naively think that four-point amplitudes are sufficient for this purpose, however if we have an invariant vertex then all amplitudes that can arise due to this interaction should satisfy the soft theorem: consistency of the four-point amplitude is a necessary but not sufficient condition. Indeed, it can be the case that vertices that are not invariant generate low-point amplitudes that satisfy our soft theorems, and the lack of invariance is only apparent for higher-point scattering where soft theorems are violated. As an example, consider the three vertices that each have only a single derivative per field: $\dot{\pi}^4$, $\dot{\pi}^2 (\partial_i \pi)^2$ and $(\partial_i \pi)^4$. Only the first of these is invariant under the $\delta \pi = b_{i}x^{i}$ symmetry, but if we take the general four-point amplitude from these interactions, with three arbitrary coupling constants, the soft theorem associated with this symmetry imposes only a single constraint (it forbids the $\dot{\pi}^2 (\partial_i \pi)^2$ vertex). We have to go to six-point scattering to also rule out the $(\partial_i \pi)^4$ vertex. We therefore have to go beyond four-point scattering if we are to use our soft theorems to search for invariant quartic vertices.

Since we are interested in field-independent symmetries, there cannot be cancellations between different topologies in a given amplitude. Furthermore, if we only have quartic vertices then all propagators are $\mathcal{O}(1)$ in the soft limit. Given that the momentum we will take soft is on-shell, requiring compatibility with all soft theorems, for all $n$-point scattering amplitudes, is equivalent to requiring that four-point amplitudes with one on-shell and three off-shell momenta satisfy the soft theorems. We would therefore like to look for solutions to our soft theorems for such semi-on-shell amplitudes and we will go through this procedure explicitly for amplitudes that scale as $\mathcal{A} \sim p^4, p^5, p^6$ i.e. that come from vertices with four, five or six derivatives, while for $\mathcal{A} \sim p^7, p^8$ we use this procedure to count the number of invariant vertices (and number of on-shell quartic amplitudes). To construct such off-shell amplitudes, we ensure that they are $SO(3)$ invariant and satisfy the leading Adler zero condition which requires each momentum to appear at least once in each term. Since we are considering off-shell amplitudes we can independently use $E$ and $\vec{p} \cdot \vec{p}$, however since we will take $p_{1} \rightarrow 0$, we will ultimately set $p_{1}^2 = 0$ and therefore $ E_1^2 = \vec{p}_1 \cdot \vec{p}_1$. 

Let's begin with amplitudes that scale as $\mathcal{A}_4 \sim p^4$. There are three off-shell amplitudes that we can write down and there are no degeneracies between these once we impose energy and momentum conservation. We write these amplitudes as
\begin{align}
\mathcal{A}_{4} = \sum_{m=1}^{3} g_{m} \mathcal{A}^{(m)}_{4}\,,
\end{align}
where 
\begin{align}
\mathcal{A}^{(1)}_4&=(\vec{p}_1\cdot \vec{p}_2) (\vec{p}_3\cdot \vec{p}_4) + \text{perms} \,, \\
\mathcal{A}^{(2)}_4&=E_3 E_4(\vec{p}_1\cdot \vec{p}_2)  + \text{perms}\,,\\
\mathcal{A}^{(3)}_4&=E_1 E_2 E_3 E_4  + \text{perms}\,.
  \end{align}
We now take $ E_1^2 = \vec{p}_1 \cdot \vec{p}_1$, impose energy and momentum conservation, and constrain this system of three amplitudes using various soft theorems.

\paragraph{$\delta \pi = c t$ symmetry} We begin with the simplest linear-in-time symmetry where we need the amplitudes to satisfy 
\begin{align} \label{SoftBoot1}
\lim_{\vec{p}_{1} \rightarrow 0} \biggl\{\frac{\tilde{\mathcal{A}}_{E_{1}}-\tilde{\mathcal{A}}_{-E_{1}}}{E_{1}} \biggr\} = 0\,.
\end{align}
We find that only a single amplitude passes this condition and corresponds to 
\begin{align}
\tilde{\mathcal{A}}_{4} = g_{1} \tilde{\mathcal{A}}^{(1)}_{4}\,.
\end{align}
In Appendix \ref{app:cosets} we go through the coset construction for this symmetry and find that there are no Wess-Zumino terms, while invariant building blocks are $\partial_i \pi$ and $\ddot{\pi}$. Given that each field must have at least one derivative acting on it and that here we are searching for vertices with exactly four derivatives, there is only a single allowed vertex which is $(\partial_i \pi)^4$. This agrees with what we found from the amplitude analysis. 

\paragraph{$\delta \pi = b_{i}x^{i}$ symmetry} Now consider the simplest purely-spatial symmetry where we need to satisfy
\begin{align} \label{SoftBoot2}
\lim_{\vec{p}_{1} \rightarrow 0} \biggl\{ \partial_{p_{i}}\left[\tilde{\mathcal{A}}_{E_{1}} + \tilde{\mathcal{A}}_{-E_{1}} \right] \biggr\} = 0\,.
\end{align}
We find that only a single amplitude passes this condition and corresponds to 
\begin{align}
\tilde{\mathcal{A}}_{4} = g_{3} \tilde{\mathcal{A}}^{(3)}_{4}\,.
\end{align}
In Appendix \ref{app:cosets} we go through the coset construction for this symmetry and find that there are no Wess-Zumino terms that have exactly four derivatives, while invariant building blocks are $\dot{\pi}$ and $\partial_{i}\partial_{j} \pi$. Given that each field must have at least one derivative acting on it and that here we are searching for vertices with exactly four derivatives, there is only a single allowed vertex which is $\dot{\pi}^4$. This agrees with what we found from the amplitude analysis. This amplitude also satisfies all the soft theorems in \eqref{Leading} and indeed the above invariant vertex has all such symmetries. 

\paragraph{$\delta \pi = c t + b_{i} x^{i}$ symmetry} In order for a vertex to be invariant under higher-order linear-in-time symmetries, it would need to be invariant under the two we have just discussed by the tower structure. However, we see that there are no common solutions to the two cases we have just discussed which suggests that there are no invariant quartic vertices which have four derivatives and this symmetry. We consider the coset construction for this symmetry in Appendix \ref{app:cosets} and find that invariant building blocks have at least two derivatives and so these cannot give rise to quartic vertices with four derivatives, while there are no Wess-Zumino terms with exactly four derivatives. The Lagrangian and amplitude analyses therefore agree. 

We now turn our attention to amplitudes that scale as $\mathcal{A}_4 \sim p^5$. There are seven off-shell amplitudes we can construct, but only two of them are non-degenerate once we impose energy and momentum conservation. We choose to write these as
\begin{align}
\mathcal{A}_{4} = \sum_{m=1}^{2} g_{m}\mathcal{A}^{(m)}_{4}\,,
\end{align}
where 
\begin{align}
\mathcal{A}^{(1)}_4&=E_1(\vec{p}_{2} \cdot \vec{p}_{2}) (\vec{p}_3\cdot \vec{p}_4) + \text{perms} \,,\\
\mathcal{A}^{(2)}_4&=E_1 E_2 E_3 \vec{p}_4 \cdot \vec{p}_4  + \text{perms} \,.
  \end{align}

\paragraph{$\delta \pi = c t$ symmetry} We again start with the simplest linear-in-time symmetry, and by demanding that the amplitude satisfies \eqref{SoftBoot1} we find that there are no solutions. As we mentioned before, there are no Wess-Zumino terms for this symmetry while invariants are built out of $\partial_i \pi$ and $\ddot{\pi}$. The only such term that we could construct with five derivatives is $(\partial_i \pi)^2 \partial_j \pi \partial_j \dot{\pi}$, however this is a total derivative and therefore yields vanishing amplitudes once energy and momentum conservation have been imposed. The Lagrangian and amplitude analyses therefore agree.

\paragraph{$\delta \pi = b_{i} x^{i}$ symmetry} For this symmetry we now demand that the amplitudes satisfy \eqref{SoftBoot2}. We find a single solution given by 
\begin{align}
\mathcal{A}_{4} = g_{2} \mathcal{A}^{(2)}_{4}\,.
\end{align}
As we show in Appendix \ref{app:cosets}, there are no Wess-Zumino terms associated with this symmetry that have exactly five derivatives, while using the invariant building blocks we find a single vertex given by $\dot{\pi}^3 \partial_i^2 \pi$. We also find that the corresponding four-point amplitude vanishes on-shell which is consistent with what we have found from the coset construction since once we go on-shell this vertex is a total derivative ($\dot{\pi}^3 \partial_i^2 \pi \sim \dot{\pi}^3 \ddot{\pi} \sim \partial_{t}(\dot{\pi}^4)$). The first non-trivial amplitude arising from this vertex is the six-point one and this amplitude has no singularities which makes sense since we can perform a field redefinition to turn this quartic vertex into a sextic one and then the six-point amplitude comes from a contact diagram rather than an exchange. This invariant vertex also has all purely-spatial symmetries and satisfies all the corresponding soft theormes as long as the symmetry parameter is traceless. We therefore again have agreement between the on-shell and off-shell analyses. We note a crucial difference between this vertex, $\dot{\pi}^3 \partial_i^2 \pi$, and the one we discussed in regards to the $\delta \pi = c t$ symmetry, $(\partial_i \pi)^2 \partial_j \pi \partial_j \dot{\pi}$. The latter is a total derivative even off-shell and so all amplitudes vanish, while the former is only a total derivative on-shell so as long as we have a Feynman diagram where at least one of the four legs is off-shell, we will get a non-zero amplitude (with the first being at six points).    

We now turn our attention to amplitudes that scale as $\mathcal{A}_4 \sim p^6$. We find 27 different possible off-shell amplitudes but a number of these are degenerate once we impose energy and momentum conservation. We are then left with 12 independent, off-shell amplitudes, and so we write 
\begin{align}
\mathcal{A}_{4} = \sum_{m=1}^{12} g_{m} \mathcal{A}^{(m)}_{4}\,,
\end{align}
where
\begin{align}
\mathcal{A}^{(1)}_4&=(\vec{p}_{1} \cdot \vec{p}_{1}) (\vec{p}_2\cdot\vec{p}_3) ( \vec{p}_2\cdot\vec{p}_4) + \text{perms}\,,\\   
\mathcal{A}^{(2)}_4&=(\vec{p}_{1} \cdot \vec{p}_{1}) (\vec{p}_{2} \cdot \vec{p}_{2})(\vec{p}_3\cdot\vec{p}_4)  + \text{perms}\,,\\
\mathcal{A}^{(3)}_4&=(\vec{p}_1\cdot\vec{p}_2) ( \vec{p}_1\cdot\vec{p}_3) ( \vec{p}_1\cdot\vec{p}_4) + \text{perms}\,,\\
\mathcal{A}^{(4)}_4&=E_1 E_3 (\vec{p}_1\cdot\vec{p}_2) ( \vec{p}_4\cdot\vec{p}_2) + \text{perms}\,,\\
\mathcal{A}^{(5)}_4&=E_1 E_3 (\vec{p}_{2} \cdot \vec{p}_{2}) (\vec{p}_1\cdot\vec{p}_4)  + \text{perms}\,,\\
\mathcal{A}^{(6)}_4&=E_1 E_2 (\vec{p}_1\cdot\vec{p}_3) ( \vec{p}_1\cdot\vec{p}_4) + \text{perms}\,,\\
\mathcal{A}^{(7)}_4&=E_1 E_2 (\vec{p}_{1} \cdot \vec{p}_{1}) (\vec{p}_3\cdot\vec{p}_4)  + \text{perms}\,,\\
\mathcal{A}^{(8)}_4&=E_2 E_3 (\vec{p}_{1} \cdot \vec{p}_{1}) (\vec{p}_1\cdot\vec{p}_4)  + \text{perms}\,,\\
\mathcal{A}^{(9)}_4&=E_1^2 E_2^2 (\vec{p}_3\cdot\vec{p}_4)  + \text{perms}\,,\\
\mathcal{A}^{(10)}_{4}&=E_1^3 E_2 (\vec{p}_3\cdot\vec{p}_4)  + \text{perms}\,,\\
\mathcal{A}^{(11)}_{4}&=E_1^2 E_2 E_3 (\vec{p}_1\cdot\vec{p}_4)  + \text{perms}\,,\\
\mathcal{A}^{(12)}_{4}&=E_1^2 E_2^2 E_3 E_4   + \text{perms}\,.
  \end{align}  
We now take $ E_1^2 = \vec{p}_1 \cdot \vec{p}_1$, impose energy and momentum conservation, and constrain this system of 12 amplitudes using various soft theorems. 

\paragraph{$\delta \pi = c t$ symmetry} We again start with the simplest linear-in-time symmetry. By demanding that the amplitude satisfies \eqref{SoftBoot1} we find a sum of seven amplitudes given by
 \begin{align}
\tilde{\mathcal{A}}_{6} = g_{1}\tilde{\mathcal{A}}^{(1)}_{6}+g_{2}\tilde{\mathcal{A}}^{(2)}_{6}+g_{3}\tilde{\mathcal{A}}^{(3)}_{6}+g_{5}(\tilde{\mathcal{A}}^{(5)}_{6} -\tilde{\mathcal{A}}^{(4)}_{6})+g_{6}(\tilde{\mathcal{A}}^{(6)}_{6} + 2\tilde{\mathcal{A}}^{(4)}_{6})+g_{7}(\tilde{\mathcal{A}}^{(7)}_{6} + 2\tilde{\mathcal{A}}^{(4)}_{6})+g_{9}\tilde{\mathcal{A}}^{(9)}_{6}\,,
\end{align}
 which suggests that there are seven invariant quartic interactions with six derivatives. As we show in Appendix \ref{app:cosets}, for this symmetry there are no Wess-Zumino terms. This means that all invariant quartic interactions need to be constructed out of $\partial_i \pi$, $\ddot{\pi}$, and additional derivatives. We find exactly seven vertices that are built out of these building blocks, have exactly six derivatives, and are not degenerate after integration by parts. These vertices are (organised in terms of the number of time derivatives):
 \begin{align}
&\partial_i \pi \partial_i \pi (\partial_j^2 \pi)^2, \qquad \partial_i \pi \partial_i \pi (\partial_j \partial_k \pi)^2, \qquad \partial_i \pi \partial_j \pi  \partial_i \partial_j \pi  \partial_k^2 \pi, \\
&\partial_i \pi \partial_i \pi  \partial_j \dot{\pi} \partial_j \dot{\pi}, \qquad \partial_i \pi \partial_i \pi (\partial_j^2 \pi) \ddot{\pi}, \qquad \partial_i \pi \partial_j \pi \partial_i \partial_j \pi \ddot{\pi},  \\ 
& \partial_i \pi \partial_i \pi \ddot{\pi}^2\,.
 \end{align}
 This counting therefore agrees with our amplitude analysis. 

 \paragraph{$\delta \pi = b_{i}x^{i}$ symmetry} Now consider the simplest purely-spatial symmetry. By demanding that we satisfy \eqref{SoftBoot2} we find a sum of five amplitudes given by
  \begin{align}
\tilde{\mathcal{A}}_{6} = g_{3}(\tilde{\mathcal{A}}^{(3)}_{6} + 3 \tilde{\mathcal{A}}^{(1)}_{6} )+g_{5}(\tilde{\mathcal{A}}^{(5)}_{6} -\tilde{\mathcal{A}}^{(4)}_{6})+g_{8}(\tilde{\mathcal{A}}^{(8)}_{6} + 2\tilde{\mathcal{A}}^{(4)}_{6})+g_{9}(\tilde{\mathcal{A}}^{(9)}_{6} + \tilde{\mathcal{A}}^{(10)}_{6}-\tilde{\mathcal{A}}^{(11)}_{6})+g_{12}\tilde{\mathcal{A}}^{(12)}_{6}\,,
\end{align}
which suggests that there are five invariant quartic vertices. As we show in Appendix \ref{app:cosets}, for this symmetry we find a single Wess-Zumino term with six derivatives, while all other invariant vertices are built out of $\dot{\pi}$ and $\partial_{i}\partial_{j} \pi$, and additional derivatives. We then find exactly four vertices that are built out of these building blocks, have exactly 6 derivatives, and are not degenerate after integration by parts. The Wess-Zumino term is
\begin{align}
(\partial_i \pi)^2 \left[(\partial_j^2 \pi)^2 - (\partial_i \partial_j \pi)^2 \right]\,,
 \end{align}
while the remaining four vertices are (organised in terms of the number of time derivatives)
\begin{align}
&\dot{\pi}^2 (\partial_j^2 \pi)^2, \qquad \dot{\pi}^2 (\partial_j \partial_k \pi)^2, \\
&\dot{\pi}^2 (\partial_j \dot{\pi})^2, \\
&\dot{\pi}^2 \ddot{\pi}^2\,.
 \end{align}
 This therefore agrees with our amplitude analysis.
 
\paragraph{$\delta \pi = c t + b_{i}x^{i}$ symmetry} Let's now combine the previous two symmetries such that we now need to satisfy both of the above soft theorems. We therefore need to pick out the common solutions of which there are two given by  
  \begin{align}
\tilde{\mathcal{A}}_{6} = g_{3}(\tilde{\mathcal{A}}^{(3)}_{6} + 3 \tilde{\mathcal{A}}^{(1)}_{6} )+g_{5}(\tilde{\mathcal{A}}^{(5)}_{6} -\tilde{\mathcal{A}}^{(4)}_{6})\,,
\end{align}
which suggests that there are two invariant quartic vertices. In Appendix \ref{app:cosets} we go through the coset construction for this symmetry and find that the invariant building blocks have at least two derivatives and therefore cannot yield quartic interactions with six derivatives. We also find two Wess-Zumino terms with six derivatives which are given by
\begin{align}
(\partial_i \pi)^2 \left[(\partial_j^2 \pi)^2 - (\partial_i \partial_j \pi)^2 \right], \qquad 
(\partial_\mu \pi)^2 \left[(\Box \pi)^2 - (\partial_\nu \partial_\rho \pi)^2 \right]\,,
 \end{align}
 which again agrees with the amplitude analysis. The first of these Wess-Zumino terms is the one we found for the $\delta \pi = b_{i} x^{i}$ symmetry and it clearly also has the $\delta \pi  = c t$ symmetry, while the second is the familiar quartic Wess-Zumino term of the Lorentz-invariant Galileon \cite{Goon:2012dy}. Note that we did not impose Lorentz invariance of the amplitude, rather imposing the soft theorem organises one of the solutions into something that is Lorentz invariant. 

 \paragraph{$\delta \pi = c_{i} t x^{i}$ symmetry} Now consider the simplest mixed symmetry. Given that our soft theorems realise the tower structure, any solutions to the full set given by \eqref{Leadingt1New} and \eqref{SubLeadingtNew} must be a sub-set of the two solutions we have just found since this symmetry is only consistent if we also have the $\delta \pi = c t + b_{i}x^{i}$ symmetry. The new condition that we must satisfy is  
  \begin{align}
\lim_{\vec{p}_{1} \rightarrow 0} \biggl\{\partial_{p_{i}} \left[ \frac{\tilde{\mathcal{A}}_{E_{1}}- \tilde{\mathcal{A}}_{-E_{1}}}{E_{1}} \right] \biggr\} = 0\,,
\end{align}
and this reduces us to a single amplitude given by
  \begin{align}
\tilde{\mathcal{A}}_{6} = g_{3}(\tilde{\mathcal{A}}^{(3)}_{6} + 3 \tilde{\mathcal{A}}^{(1)}_{6} )\,,
\end{align}
which suggests that there is only a single invariant quartic vertex with six derivatives. In Appendix \ref{app:cosets} we go through the coset construction for this symmetry and as above the invariant building blocks have at least two derivatives and therefore cannot yield quartic interactions with six derivatives. We then find a single Wess-Zumino term given by 
\begin{align}
(\partial_i \pi)^2 \left[(\partial_j^2 \pi)^2 - (\partial_i \partial_j \pi)^2 \right]\,.
 \end{align}
We therefore again agree with the amplitude analysis. 

 \paragraph{$\delta \pi = b_{i_{1} \ldots i_{n}} x^{i_{1}} \ldots x^{i_{n}}$ symmetry} Now consider the general purely-spatial symmetry meaning that we need to satisfy all soft theorems in \eqref{Leading}. We find three solutions given by 
   \begin{align}
\tilde{\mathcal{A}}_{6} = g_{8}(\tilde{\mathcal{A}}^{(8)}_{6} + 2\tilde{\mathcal{A}}^{(4)}_{6})+g_{9}(\tilde{\mathcal{A}}^{(9)}_{6} + \tilde{\mathcal{A}}^{(10)}_{6}-\tilde{\mathcal{A}}^{(11)}_{6})+g_{12}\tilde{\mathcal{A}}^{(12)}_{6}\,,
\end{align}
and so we expect three invariant quartic vertices. Given that we take the symmetry parameters to be traceless we have two possible building blocks which are $\dot{\pi}$ and $\partial_i^2 \pi$ plus additional derivatives, and indeed we find three invariant vertices that have exactly six derivatives and are not degenerate after integration by parts. These are
\begin{align}
\dot{\pi}^2 \ddot{\pi}^2, \qquad \dot{\pi}^2 \ddot{\pi} \partial_i^2 \pi, \qquad \dot{\pi}^2 (\partial_i^2 \pi)^2\,.
 \end{align}
We therefore again agree with the amplitude analysis. 

 \paragraph{$\delta \pi = c t +  b_{ij} x^{i} x^{j}$ symmetry} As a final example consider the simplest linear-in-time symmetry plus the quadratic purely-spatial symmetry. By the tower structure, any solutions to the corresponding soft theorems must be contained in the solutions corresponding to the $\delta \pi = c t + b_{i}x^{i} $ symmetry of which we found two. If we now impose the additional condition here which is
   \begin{align}
b_{ij}\lim_{\vec{p}_{1} \rightarrow 0} \biggl\{\partial_{p_{i}} \partial_{p_{j}}  \left[\tilde{\mathcal{A}}_{E_{1}} + \tilde{\mathcal{A}}_{-E_{1}} \right] \biggr\} = 0\,,
\end{align}
 then we find no solutions and therefore there shouldn't be any invariant quartic vertices with six derivatives. Indeed, it is straightforward to check that the Wess-Zumino terms that we found for the $\delta \pi = c t + b_{i}x^{i} $ symmetry are not invariant under the $\delta \pi = b_{ij} x^{i} x^{j}$ symmetry so this agrees with the amplitude analysis. Given that there are no solutions in this case, the tower structure ensures that there are also no solutions if we include additional powers of $x^i$ in either the $c t$ or $b_{ij} x^{i} x^{j}$ part of the symmetry. 

 So far we have been focused on off-shell four-point amplitudes since we wanted to find invariant quartic interactions for which our soft theorems should be satisfied by all corresponding $n$-point amplitudes. We argued that this requires the semi-on-shell four-point amplitude to satisfy the soft theorems. We went through this procedure for a number of different momentum scalings and a number of different symmetries and in all cases the solutions to our soft theorems match what we get from building invariant Lagrangians using the coset construction. However, the soft bootstrap procedure is of course usually concerned with building on-shell amplitudes, and in the on-shell limit a number of off-shell amplitudes we have constructed are degenerate. In Tables \ref{tab:p4} - \ref{tab:p8} we present the number of off-shell and on-shell amplitudes that satisfy our soft theorems for various symmetries. In all cases ``$\ldots$" indicates the possible inclusion of the purely-spatial symmetries. We also include the counting for amplitudes scaling as $\tilde{\mathcal{A}}_4 \sim p^7, p^8$.
\begin{table} 
\begin{center}
\begin{tabular}{| c | c | c | c | c |} 
 \hline
 Symmetries & $1$ & $t$ & $x$ & $t+x+\ldots$\\ [0.5ex] 
 \hline
 off-shell $\mathcal{A}_4 \sim p^4$ & 3 & 1 & 1 & 0\\ 
 \hline
 on-shell $\mathcal{A}_4 \sim p^4$ & 3 & 1 & 1 & 0 \\
 \hline
\end{tabular}
\caption{Number of four-point amplitudes with a $\mathcal{A}_{4} \sim p^4$ scaling}\label{tab:p4}
\end{center}
\begin{center}
\begin{tabular}{| c | c | c | c |} 
 \hline
 Symmetries & $1$ & $t+\ldots$ & $x+\ldots$ \\ [0.5ex] 
 \hline
 off-shell $\mathcal{A}_4 \sim p^5$ & 2 & 0 & 1 \\ 
 \hline
 on-shell $\mathcal{A}_4 \sim p^5$ & 0 & 0 & 0  \\
 \hline
\end{tabular}
 \caption{Number of four-point amplitudes with a $\mathcal{A}_{4} \sim p^5$ scaling}
\end{center}
\begin{center}
\begin{tabular}{| c | c | c | c | c | c | c | c |} 
 \hline
 Symmetries & $1$ & $t$ & $x$ & $t+x$ & $tx$ & $xx+\ldots$ & $t+xx+\ldots$\\ [0.5ex] 
 \hline
 off-shell $\mathcal{A}_4 \sim p^6$ & 12 & 7 & 5 & 2 & 1 & 3 & 0\\ 
 \hline
 on-shell $\mathcal{A}_4 \sim p^6$ & 6 & 4 & 3 &  2 & 1 & 1 & 0\\
 \hline
\end{tabular}
 \caption{Number of four-point amplitudes with a $\mathcal{A}_{4} \sim p^6$ scaling}
\end{center}
\begin{center}
\begin{tabular}{| c | c | c | c | c | c | c | c |} 
 \hline
 Symmetries & $1$ & $t$ & $x$ & $t+x+\ldots$ & $xx+\ldots$ \\ [0.5ex] 
 \hline
 off-shell $\mathcal{A}_4 \sim p^7$ & 15 & 7 & 8 & 0 & 6 \\ 
 \hline
 on-shell $\mathcal{A}_4 \sim p^7$ & 3 & 2 & 2 &  0 & 1 \\
 \hline
\end{tabular}
 \caption{Number of four-point amplitudes with a $\mathcal{A}_{4} \sim p^7$ scaling}
\end{center}
\begin{center}
\begin{tabular}{| c | c | c | c | c | c | c | c | c | c | c |} 
 \hline
 Symmetries & $1$ & $t$ & $x$ & $t+x$ & $tx$ & $xx$ & $t+xx\ldots$ & $tx+xx\ldots$ & $x^3\ldots$ \\ [0.5ex] 
 \hline
 off-shell $\mathcal{A}_4 \sim p^8$ & 44 & 31 & 26 & 18 & 11 & 17 & 9 & 5 & 16\\ 
 \hline
 on-shell $\mathcal{A}_4 \sim p^8$  & 13 & 10 & 9 & 8 &  4 & 5 & 3 & 1 & 4\\
 \hline
\end{tabular}
 \caption{Number of four-point amplitudes with a $\mathcal{A}_{4} \sim p^8$ scaling} \label{tab:p8}
\end{center}
\end{table}

So far in this section our focus has been on quartic vertices but let us also put our soft theorems to the test by constructing six-point interactions. Here we restrict ourselves to those with six derivatives, and again we consider semi-on-shell six-point amplitudes as a means to construct interactions for which all $n$-point on-shell amplitudes satisfy the soft theorems. Again, the only leg that we take on-shell is the one we will ultimately take soft. At sextic order in momenta there are four off-shell amplitudes that we can construct and they are all independent after imposing energy and momentum conservation (recall that we take each momentum to appear at least once in order to satisfy the leading Adler zero condition). We have
\begin{align}
\mathcal{A}_{6} = \sum_{m=1}^{4} g_m \mathcal{A}^{(m)}_{6}\,,
\end{align}
 where 
\begin{align}
\mathcal{A}^{(1)}_6&=(\vec{p}_1\cdot \vec{p}_2) (\vec{p}_3\cdot \vec{p}_4) (\vec{p}_5\cdot \vec{p}_6) + \text{perms} \,, \\
 \mathcal{A}^{(2)}_6&=E_5 E_6(\vec{p}_1\cdot \vec{p}_2) (\vec{p}_3\cdot \vec{p}_4)  + \text{perms}\,,\\
\mathcal{A}^{(3)}_6&=E_3 E_4 E_5 E_6(\vec{p}_1\cdot \vec{p}_2)  + \text{perms}\,, \\
\mathcal{A}^{(4)}_6&=E_1 E_2 E_3 E_4 E_5 E_6  + \text{perms}\,.
  \end{align}
As before we now set $ E_1^2 = \vec{p}_1 \cdot \vec{p}_1$ and impose various soft theorem conditions. If we consider the $\delta \pi = b_{i} x^{i}$ symmetry then we find that only the $g_{4}$ term satisfies the soft theorem and indeed it satisfies all purely-spatial soft theorems. The corresponding interaction is $\dot{\pi}^6$. While if we consider the $\delta \pi = c t$ symmetry then only the $g_{1}$ term is allowed and it doesn't satisfy any other soft theorems. The corresponding interaction is $(\partial_i \pi)^6$. These results are summarised in Table \ref{tab:a6}.
\begin{table}
\begin{center}
\begin{tabular}{| c | c | c | c | c |} 
 \hline
 Symmetries & $1$ & $t$ & $x$ & $t+x+\ldots$\\ [0.5ex] 
 \hline
 off-shell $\mathcal{A}_6 \sim p^6$ & 4 & 1 & 1 & 0\\ 
 \hline
 on-shell $\mathcal{A}_6 \sim p^6$ & 4 & 1 & 1 & 0 \\
 \hline
\end{tabular}
\caption{Number of six-point amplitudes with a $\mathcal{A}_{6} \sim p^6$ scaling}\label{tab:a6}
\end{center}
\end{table}
Generalising to higher-point interactions and amplitudes is somewhat tedious but straightforward. 


\section{Conclusion and outlook} \label{sec:conclusion}

In this paper we have derived soft theorems that must be satisfied by on-shell amplitudes if the underlying theory has a non-linearly realised symmetry. Motivated by cosmology and cosmological correlators, we considered boost-breaking amplitudes and non-linear symmetries that treat time and space on different footings. Such amplitudes are contained within cosmological correlators in a particular singular limit and so having a solid understanding of the connection between cosmological observables and non-linear symmetries requires a solid understanding of such boost-breaking amplitudes. We derived these soft theorems using current conservation and the LSZ reduction formula, and by being careful with derivatives and delta functions, we realised a tower of soft theorems which is ultimately related to closure of the underlying symmetry algebra. Our soft theorems require combination of amplitudes to vanish with the rate at which they vanish dictated by the structure of the corresponding symmetry. We considered general soft solutions to these soft theorems and checked that we recover known Lorentz-invariant results. We further assessed the validity of our soft theorems by using them to construct \textit{off-shell} invariant quartic vertices and found agreement in a large number of examples with what we find using the coset construction.

There are a number of avenues for future work:

 \begin{itemize}
     \item The soft theorems we derived in this paper take the linear dispersion relation $E_{p} = |\vec{p}|$ as an input. To match with off-shell examples, we find that our soft theorems demand symmetry parameters to be traceless and this is ultimately due to the fact we impose this linear dispersion relation. Indeed, the two-derivative kinetic term is only invariant if the symmetry parameters are traceless. It would be natural to extend the scope of our work to include non-linear dispersion relations that are discussed in e.g. \cite{Griffin:2014bta}. We actually expect that the form of our soft theorems will not change as the dispersion relation changes, however what we mean by $E$ in our soft theorems will indeed change and therefore the solutions could impose other conditions on the symmetry parameters (for example, vanishing of the double trace). 
    \item We imposed a  $\pi \rightarrow -\pi$ symmetry in order to exclude quadratic terms in the current which can introduce poles in the soft amplitudes. It would be interesting to extend our analysis to include such terms by relaxing the $\pi \rightarrow -\pi$ symmetry and we will explore this avenue in an upcoming paper \cite{PeterDavid2}.
    \item As we mentioned in Section \ref{sec:derivation}, the structure of some of our soft theorems resembles the shifted wavefunction coefficient that appears in the soft theorems for the flat-space wavefunction in \cite{Bittermann:2022nfh}. It would be interesting to investigate this connection further and to understand how the tower structure arises for wavefunction soft theorems. It would then be natural to extend the analysis to the inflationary wavefunction. 
 \end{itemize}

 \paragraph{Acknowledgements} We thank Clifford Cheung, Trevor Cheung, Sadra Jazayeri, Austin Joyce, Xi Tong, Yuhang Zhu and Shuang-Yong Zhou for helpful discussions. D.S. is supported by a UKRI Stephen Hawking
Fellowship [grant number EP/W005441/1] and a Nottingham Research Fellowship from the University
of Nottingham. ZD is supported by Nottingham CSC [file No. 202206340007].
For the purpose of open access, the authors have applied a CC BY public copyright licence to any Author
Accepted Manuscript version arising. 

\paragraph{Data access statement} No new data were
created or analysed during this study. 


\appendix

\section{Coset constructions and Wess-Zumino terms} \label{app:cosets}

In this appendix we go through the coset construction for a number of examples and construct quartic Wess-Zumino terms, with a particular focus on those that yield amplitudes with a $p^{6}$ scaling so that we can compare with what we found in Section \ref{sec:bootstrap}. We focus on the non-linear symmetries: $\delta \pi = c t, b_i x^i, c t + b_i x^i, c_i t x^i $, and in all cases we assume that the linearly realised symmetries are spacetime translations (generated by $P_{0}$ and $P_{i}$) and spatial rotations (generated by $J_{ij}$). We therefore always have an unbroken subgroup with the following non-zero commutators:
\begin{align} 
        [P_i, J_{jk}] &= -\delta_{ij}P_k + \delta_{ik}P_{j}\\
        [J_{ij},J_{kl}] &= -\delta_{ik}J_{jl} + \delta_{il}J_{jk} + \delta_{jk}J_{il} - \delta_{jl}J_{ik}\,.
\end{align}
We also assume a constant shift symmetry in each case which is generated by $A$. Since $A$ is a $SO(3)$ scalar and generates a constant shift symmetry, it commutes with the generators of the unbroken subgroup. It also commutes with all other generators that we might add due to the tower structure which was explained in detail in \cite{Roest:2019oiw,GJS}. The idea is that the Goldstone mode associated with this generator is $\pi$, which is the field we are interested in constructing theories of. We are then interested in additional symmetries that are non-linearly realised by $\pi$. This requires us to add additional broken generators in the coset construction which in turn requires us to introduce additional fields in the coset element. We then rely on \textit{inverse Higgs constraints} to eliminate these additional fields. The existence of inverse Higgs constraints, which enable us to algebraically solve for these additional fields in terms of $\pi$ and its derivatives, requires there to be a certain structure in the commutators: the commutator between a broken generator and a translation (space or time) must yield another broken generator. This naturally yields a tower structure in which successive actions of translations should land us on $A$ (paths that do not lead to $A$ are ruled out by Jacobi identities \cite{Roest:2019oiw}). For example, if we have the commutator $[P_{0}, C]  = A$ then we will have an inverse Higgs constraint of the form $c \sim \dot{a}$, while the commutator $[P_{i}, B_{j}]  = \delta_{ij} A$ will yield $b_{i} \sim \partial_{i} a$. If we also have $[P_{0}, C']  = C$, then we will have $c' \sim \dot{c} \sim \ddot{a}$.

\paragraph{$\delta \pi = c \, t$ symmetry} In this case we have a single additional generator which is $C$. The additional non-zero commutator is
\begin{align}
        [P_0, C] =  A\,. \label{tTowerGenerator}
\end{align}
To construct the invariant building blocks of invariant actions, and to construct Wess-Zumino terms, we need to compute the Maurer-Cartan form from the coset element $g$. Choosing the parametrisation $g=e^{x^i P_i}e^{t P_0}e^{\pi A}e^{\xi_0 C}$, yields 
\begin{equation}
    \omega = g^{-1} dg = \omega_{P}^i P_i + \omega_{P}^0 P_0 + \omega_{A} A + \omega_{C} C \,,
\end{equation}
where 
\begin{align}
    \omega_{P}^i &= dx^i\,, \nonumber\\
    \omega_{P}^0 &= dt\,, \nonumber\\
    \omega_{A} &= d\pi + \xi_0 dt\,, \nonumber\\
    \omega_C &= d\xi_0 \,.\label{tOneForms}
\end{align}
Since $\omega_{A}$ transforms covariantly under all symmetries, it is consistent to set it to zero. After pulling back to spacetime, this yields the inverse Higgs constraint: $\xi_0 = -\dot\pi$. The invariant building blocks then come from inserting this solution back into the above one-forms and pulling back to spacetime. From $\omega_A$ we then have $\partial_{i}\pi$ as a building block, while from $\omega_C$ we have $\partial_{i} \dot{\pi}$ and $\ddot{\pi}$. Since we are free to add additional derivatives to these building blocks the invariant building blocks are simply $\partial_{i}\pi$ and $\ddot{\pi}$.

Wess-Zumino terms do not follow from this procedure. Rather, to compute them we need to build closed five-forms out of the above one-forms. Writing such a five-form as $\omega_5 = d \omega_4$, Wess-Zumino terms follow from the pull back of $\omega_4$. This is explained in detail in \cite{Goon:2012dy} so we refer the reader there for more details. Since each one-form associated with the broken generators is exactly linear in $\pi$ (once we impose the inverse Higgs constraints), an $n$-point self-interaction would require the five-form to contain $n$ copies of the one-forms associated with broken generators. This immediately implies that there are no Wess-Zumino terms beyond quintic order in the fields. For this particular example, we cannot wedge together more than one copy of $\omega_{A}$ or $\omega_{C}$, and so there cannot be any Wess-Zumino terms beyond quadratic order in the fields. The only five-form that will yield something quadratic in the fields, and is consistent with rotations, is then 
\begin{align}
\omega_5 = \epsilon_{ijk} \omega_{P}^i \wedge \omega_{P}^{j} \wedge \omega_{P}^k \wedge \omega_{A} \wedge \omega_{C}\,.
\end{align}
We can easily check closure:
\begin{align}
d \omega_5 = \epsilon_{ijk} \omega_{P}^i \wedge \omega_{P}^{j} \wedge \omega_{P}^k \wedge (d \xi^{0} \wedge dt) \wedge \omega_{C} = \epsilon_{ijk} \omega_{P}^i \wedge \omega_{P}^{j} \wedge \omega_{P}^k \wedge (\omega_{C} \wedge dt) \wedge \omega_{C} = 0 \,.
\end{align}
We then have
\begin{align}
\omega_4 = -  \pi \epsilon_{ijk} \omega_{P}^i \wedge \omega_{P}^{j} \wedge \omega_{P}^k \wedge d \xi^{0} -\frac{1}{2} \xi_{0}^2 \epsilon_{ijk} d t \wedge  \omega_{P}^i \wedge \omega_{P}^{j} \wedge \omega_{P}^k \,.
\end{align}
Once we pull back to spacetime, drop the volume factor, and impose the inverse Higgs constraint, this Wess-Zumino term, after integration by parts, is then simply the kinetic term $\dot{\pi}^2$. It is simple to check that this term when appearing in the Lagrangian is only invariant under $\delta \pi = c  t$ up to a total derivative which is the tell-tale sign that this is a Wess-Zuimno term. Note that the gradient term $(\partial_i \pi)^2$ is not a Wess-Zumino term since it can be constructed out of the invariant building block $\partial_i \pi$ and is trivially invariant under $\delta \pi = c t$. At linear order in the fields the only closed five-form is 
\begin{align}
\omega_5 = \epsilon_{ijk} \omega_{P}^i \wedge \omega_{P}^{j} \wedge \omega_{P}^k \wedge \omega_{P}^{0} \wedge \omega_{C} \,,
\end{align}
and this yields the tadpole $\dot{\pi}$. If we were to replace $\omega_{C}$ with $\omega_{A}$ then the five-form would not be closed. From now on we will focus on quartic Wess-Zumino terms. 

\paragraph{$\delta \pi = b_i x^i$ symmetry} In this case we again have a single additional generator but now with two non-zero new commutators:
\begin{align}
        [P_i, B_j] &= - \delta_{ij} A \label{xTowerGenerator} \,,\\
        \;\;[J_{ij}, B_k] &= -\delta_{ik} B_j + \delta_{jk} B_i \,.
\end{align}
The Maurer-Cartan form following from the coset parametrisation $g=e^{x^i P_i}e^{t P_0}e^{\pi A}e^{\xi^i B_i}$ is
\begin{equation}
    \omega = g^{-1} dg = \omega_{P}^i P_i + \omega_{P}^0 P_0 + \omega_{A} A + \omega_{B}^i B_i\,,
\end{equation}
where 
\begin{align}
    \omega_{P}^i &= dx^i\,, \nonumber\\
    \omega_{P}^0 &= dt\,, \nonumber\\
    \omega_{A} &= d\pi + \xi_i dx^i\,, \nonumber\\
    \omega_{B}^i &= d\xi^i\,.\label{xOneForms}
\end{align} 
The story is now somewhat similar to the previous example: the inverse Higgs constraint that allow us to eliminate $\xi_i$ is simply $\xi_i = -\partial_i\pi$, and the resulting invariant building blocks are $\dot{\pi}$ and $\partial_{i} \partial_{j} \pi$. Again there is a Wess-Zumino term at quadratic order which is the gradient term $(\partial_i \pi)^2$ so let us turn our attention to quartic interactions. We therefore need to build a five-form that is quartic in $\omega_{A}$ and $\omega_{B}^i$. Clearly we can have at most one of the former, while we also cannot have more then three of the latter since when wedged together four or more will vanish (since there are only three components in $\omega_{B}^i$). The only possibility is then    
\begin{equation}
    \omega_5  = \epsilon_{ijk} \omega_{A} \wedge \omega_{B}^{i} \wedge \omega_{B}^{j} \wedge \omega_{B}^{k} \wedge \omega_{P}^{0} \,.
\end{equation}
Closure of this five-form then follows from the fact that $d \omega_{A} = \delta_{ij} \omega_{B}^{i} \wedge \omega_{P}^{j}$. By finding the corresponding $\omega_4$, pulling back to spacetime, imposing the inverse Higgs constraint and dropping the volume factor yields the self-interaction: $\pi[(\partial_{i} \partial_{i} \pi)^3 - 3\partial_{i} \partial_{i}\pi (\partial_{j} \partial_{k} \pi)^2 + 2 \partial_{i}\partial_{j} \pi \partial_{j}\partial_{k}\pi \partial_{k}\partial_{l}\pi)]$, which, unsurprisingly, takes an identical form as the Lorentz-invariant quartic Galileon Wess-Zumino term derived in \cite{Goon:2012dy}, but now with Lorentzian derivatives replaced by spatial ones. Up to a total derivative, this interaction is equivalent to $(\partial_{i} \pi)^2[(\partial_{i} \partial_{i} \pi)-(\partial_{i}\partial_{j} \pi)^2]$. This is the only quartic Wess-Zumino term for this symmetry, which is consistent with what we found using our soft theorem in Section \ref{sec:bootstrap}.

\paragraph{$\delta\pi = c t + b_i x^i$ symmetry} Let's now consider the combination of the previous two symmetries. We don't assume that the interactions we will derive are Lorentz invariant so the results are not simply those of \cite{Goon:2012dy}. The non-zero commutators involving broken generators are 
\begin{align}
        [P_i, B_j] &= - \delta_{ij} A \,, \\
        [P_0, C] &=  A \,, \\
        \;\;[J_{ij}, B_k] &= -\delta_{ik} B_j + \delta_{jk} B_i \,.
\end{align}
The Maurer-Cartan form is then
\begin{align}
        \omega = g^{-1} dg = \omega^i P_i + \omega^0 P_0 + \omega_{A} A + \omega_{C} C +\omega_{B}^{i} B_i \,,
\end{align}
where
\begin{align}
    \omega_{P}^i &= dx^i\,, \nonumber\\
    \omega_{P}^0 &= dt\,, \nonumber\\
    \omega_{A} &= d\pi + \xi_0 dt + \xi_i dx^i\,, \nonumber\\
    \omega_{C} &= d\xi^0 \,, \nonumber\\
    \omega_{B}^i &= d\xi^i \,.
\end{align}
The solutions to the inverse Higgs constraints are $\xi_0 = -\dot\pi$ and $\xi_i = -\partial_i \pi$, and the building blocks of invariant Lagrangians are $\partial_{i} \partial_{j} \pi$, $\partial_{i} \dot{\pi}$ and $\ddot{\pi}$. Turning to Wess-Zumino terms, we note that all of the one-forms are closed other than $\omega_{A}$ which satisfies $d \omega_{A} = \omega_C \wedge \omega_P^0 + \delta_{ij}\omega_{B}^{i} \wedge \omega_{P}^{j} $. There are then three closed five-forms that we can build that can yield quartic interactions for $\pi$. They are 
\begin{align}
    \omega_5^{(1)} & =  \epsilon_{ijk} \omega_{A} \wedge \omega_{B}^{i} \wedge \omega_{B}^{j} \wedge \omega_{B}^{k} \wedge \omega_{P}^{0} \,, \\
    \omega_5^{(2)} & =  \epsilon_{ijk} \omega_{A} \wedge \omega_{C} \wedge \omega_{B}^{i} \wedge \omega_{B}^{j} \wedge \omega_{P}^{k}\,, \\
    \omega_{5}^{(3)} & =  \epsilon_{ijk} \omega_{C} \wedge \omega_{B}^{i} \wedge \omega_{B}^{j} \wedge \omega_{B}^{k} \wedge \omega_{P}^{0} \,.
\end{align}
It turns out that the interaction we get from $\omega_{5}^{(3)}$ is a total derivative so we ignore it, while the second can be written as  
\begin{align}
    \tilde{\omega}_5^{(2)} = \epsilon_{\mu\nu\rho\sigma} \omega_{A} \wedge \omega_{B}^\mu \wedge \omega_{B}^\nu \wedge \omega_{B}^\rho \wedge \omega_{P}^\sigma  \,,
\end{align}
by allowing ourselves to add any amount of $\omega_5^{(1)}$ to it. Here $\omega_{B}^0 \equiv \omega_C $ and $\{ \mu\nu\rho\sigma \}$ are Lorentzian indices. By pulling back to spacetime, imposing the inverse Higgs constraints and dropping the volume factors we find the following self-interactions: $\pi[(\partial_{i}^2 \pi)^3 - 3\partial_{i}^2 \pi (\partial_{j} \partial_{k} \pi)^2 + 2 \partial_{i}\partial_{j} \pi \partial_{j}\partial_{k}\pi \partial_{k}\partial_{l}\pi)]$ and $\pi[(\partial_{\mu}^2 \pi)^3 - 3\partial_{\mu}^2 \pi (\partial_{\nu} \partial_{\rho} \pi)^2 + 2 \partial_{\mu}\partial_{\nu} \pi \partial^{\nu}\partial_{\rho}\pi \partial^{\rho}\partial^{\mu}\pi)]$ which can be written as $(\partial_i \pi)^2 \left[(\partial_j^2 \pi)^2 - (\partial_i \partial_j \pi)^2 \right]$ and $ (\partial_\mu \pi)^2 \left[(\Box \pi)^2 - (\partial_\nu \partial_\rho \pi)^2 \right]$, respectively. There are no other quartic Wess-Zumino terms which is again consistent with what we found in Section \ref{sec:bootstrap}.

\paragraph{$\delta\pi = c_i t x^i$ symmetry} This is our final example and to not cause confusion with the $C$ generator, let us write the generator of this symmetry as $S_{i}$. The non-trivial commutators are then
\begin{align}
        [P_i, B_j] &= - \delta_{ij} A \,, \\
        [P_0, C] &=  A\,, \\
        [P_{i}, S_{j}] &= -\delta_{ij} C\,, \\
        [P_0, S_{i}] &= B_{i}\,, \\
        [J_{ij}, S_{k}] &= -\delta_{ik} S_{j} +\delta_{jk} S_{i}\,, \\
        \;\;[J_{ij}, B_k] &= -\delta_{ik} B_j + \delta_{jk} B_i\,.
\end{align}
The Maurer-Cartan form is
\begin{align}
\omega = g^{-1} dg = \omega^i P_i + \omega^0 P_0 + \omega_{A} A + \omega_{C} C +\omega_{B}^{i} B_i + \omega_{S}^{i} S_{i} \,,
\end{align}
where
\begin{align}
    \omega_{P}^i &= dx^i\,, \\
    \omega_{P}^0 &= dt\,, \\
    \omega_{A} &= d\pi + \xi_0 dt + \xi_i dx^i\,,\\
    \omega_{C} &= d\xi^0 +\phi_{i} dx^{i} \,, \\
    \omega_{B}^i &= d\xi^i + \phi^{i}dt \label{Boneform} \,,\\
    \omega_{S}^i &= d\phi^{i} \,.
\end{align}
The solutions to the inverse Higgs constraints are $\xi_0 = -\dot\pi$, $\xi_i = -\partial_i \pi$ and $\phi_i = \partial_i\dot\pi$, and the invariant building blocks are $\partial_{i}\partial_j \pi$ and $\ddot\pi$. Note that compared to the previous example, the invariant building block $\partial_{i} \dot{\pi}$ is no longer allowed. To compute quartic Wess-zumino terms we note that the one-forms satisfy
\begin{align}
     d \omega_{P}^{i} &= 0 \,,\\
     d \omega_{P}^{0} &= 0 \,,\\
     d \omega_{A} &= \omega_C \wedge \omega_P^0 - \delta_{ij}\omega_{B}^{i} \wedge \omega_{P}^{j} \,,\\
     d \omega_C &= -\delta_{ij} \omega_S^i \wedge\omega_P^j \,, \\
     d \omega_{B}^i &= \omega_S^i \wedge \omega_P^0 \,,  \\
     d \omega_{S}^i &= 0 \,,
\end{align}
and using these relations we can build five closed five-forms which are
\begin{align}
    \omega_{5}^{(1)} =  &\epsilon_{ijk} \omega_{A} \wedge \omega_{B}^{i}\wedge\omega_{B}^{j}\wedge\omega_{B}^{k}\wedge \omega_{P}^{0} \,,\\
       \omega_{5}^{(2)} =     &\epsilon_{ijk} \omega_{S}^{i}\wedge\omega_{S}^{j}\wedge\omega_{S}^{k}\wedge\omega_{C} \wedge \omega_{P}^{0} \,,\\
     \omega_{5}^{(3)}=       &\epsilon_{ijk} \omega_{S}^{i}\wedge\omega_{S}^{j}\wedge\omega_{S}^l\wedge\omega_{B}^{l}\wedge \omega_{P}^{k} \,,\\
     \omega_{5}^{(4)} =       &\epsilon_{ijk} \omega_{S}^{i}\wedge\omega_{S}^{j}\wedge\omega_{S}^{k}\wedge\omega_{B}^{l}\wedge \omega_{P}^{l} \,,\\
      \omega_{5}^{(5)} =       &\epsilon_{ijk} \omega_{S}^{i}\wedge\omega_{S}^{j}\wedge\omega_{S}^{l} \wedge \omega_{B}^{k} \wedge \omega_{P}^{l}\,.
\end{align}
Given how the one forms depend on $\pi$, once we impose all inverse Higgs constraints, only the first of these can yield a quartic interaction with six derivatives. The resulting Wess-Zumino term is the one we have now seen a number of times: $(\partial_i \pi)^2 \left[(\partial_j^2 \pi)^2 - (\partial_i \partial_j \pi)^2 \right]$.  Note that is crucial here that only the coefficient of $d t$ in \eqref{Boneform} is set to zero when we eliminate the extra fields thereby allowing this first five-form to yield a non-zero result once we pull back and impose the inverse Higgs constraints. The existence of a single Wess-Zumino term with six derivatives agrees with what we found in Section \ref{sec:bootstrap}.   Moreover, the resulting vertex $(\partial_i\dot\pi)^{2}\left[ (\partial_{j}^{2}\dot\pi)^2 -(\partial_{i}\partial_{j}\dot\pi) \right] $ from pulling back $\omega_{5}^{(2)} $ precisely matches our soft theorem prediction from example 5 in Section \ref{sec:derivation}.


\bibliographystyle{JHEP}
\bibliography{refsAdlerZero}

\end{document}